\documentclass[%
 reprint, 
superscriptaddress,
 amsmath,amssymb,
 aps, physrev,
]{revtex4-2}

\usepackage{graphicx}% Include figure files
\usepackage{dcolumn}% Align table columns on decimal point
\usepackage{bm}% bold math

\usepackage{amssymb,amsmath}
\usepackage{subfigure}  % instead of subcaption
\usepackage{hyperref}
\usepackage{xcolor}

\usepackage{geometry}
\begin{document}

% Use the \preprint command to place your local institutional report
% number in the upper righthand corner of the title page in preprint mode.
% Multiple \preprint commands are allowed.
% Use the 'preprintnumbers' class option to override journal defaults
% to display numbers if necessary
%\preprint{}

%Title of paper
\title{\textbf{Scattered light noise due to dust particles contamination in the vacuum pipes of the Einstein Telescope} 
}% 
% repeat the \author .. \affiliation  etc. as needed
% \email, \thanks, \homepage, \altaffiliation all apply to the current
% author. Explanatory text should go in the []'s, actual e-mail
% address or url should go in the {}'s for \email and \homepage.
% Please use the appropriate macro foreach each type of information

% \affiliation command applies to all authors since the last
% \affiliation command. The \affiliation command should follow the
% other information
% \affiliation can be followed by \email, \homepage, \thanks as well.
\author{Andrea Moscatello}
\email{andrea.moscatello@phd.unipd.it}
\affiliation{Physics and Astronomy Department, University of Padova, Via Marzolo 8, 35131, Padova, Italy}
\affiliation{INFN, Padova Section, Via Marzolo 8, 35131 Padova, Italy}%

\author{Giacomo Ciani}
\email{giacomo.ciani@unitn.it}
\affiliation{Physics Department, University of Trento, Via Sommarive 14, 38123 Povo (TN), Italy}%
\affiliation{TIFPA - INFN, Via Sommarive, 14, 38123 Povo (TN), Italy}%

\author{Livia Conti}
\email{livia.conti@pd.infn.it}
\affiliation{INFN, Padova Section, Via Marzolo 8, 35131 Padova, Italy}%

\author{Marco Bazzan}
%\email{marco.bazzan@unipd.it}
\affiliation{Physics and Astronomy Department, University of Padova, Via Marzolo 8, 35131, Padova, Italy}
\affiliation{INFN, Padova Section, Via Marzolo 8, 35131 Padova, Italy}%

% \author{Gianpietro Cagnoli}
% \email{gianpietro.cagnoli@unipd.it}
% \affiliation{Physics and Astronomy Department, University of Padova, Via Marzolo 8, 35131, Padova, Italy}
% \affiliation{INFN, Padova Section, Via Marzolo 8, 35131 Padova, Italy}

\author{Jean-Pierre Zendri}
%\email{jean-pierre.zendri@pd.infn.it}
\affiliation{INFN, Padova Section, Via Marzolo 8, 35131 Padova, Italy}
%Collaboration name if desired (requires use of superscriptaddress
%option in \documentclass). \noaffiliation is required (may also be
%used with the \author command).
%\collaboration can be followed by \email, \homepage, \thanks as well.
%\collaboration{}
%\noaffiliation

\date{\today}

\begin{abstract}
High-sensitivity optical measurements such as those performed in interferometric Gravitational Wave detectors are prone to scattered light noise. To minimize it, optical components must meet tight requirements on surface roughness and bulk defects. Nonetheless, the effectiveness of these measures can be undermined by other sources of scattered light. In this article, we examine scattered light noise caused by particles deposited on surfaces, especially on the baffles inside the vacuum pipes of the Einstein Telescope’s interferometer arms. First, we study light scattering by particles deposited on a surface and having diameters from about one tenth to hundred times the light wavelength: we discuss its angular distribution and dependence on particle size and refractive index, and on polarization. Then, we specialize to the case of the Einstein Telescope arms and quantify the maximum allowed density of particles on each arm baffle. We conclude with cleanliness guidelines for the assembly of the vacuum pipes, including the required cleanliness class of the installation environment.
\end{abstract}

\maketitle

\section{Introduction}\label{sec:intro}
Scattered light (SL) inside the laser interferometers was one of the dominant sources of noise at low-mid frequency during the observing period O3 of the LIGO-Virgo-KAGRA Collaboration in search of Gravitational Wave (GW) signals \cite{PhysRevX.13.041039}; 
it remained a significant limiting noise source throughout the O4 observing run in the frequency band below a few hundred Hz \cite{nandi2026scatteredlightnoiseligo}. Several mechanisms allow scattered light to generate noise: either by recoupling to the main beam, and hence directly polluting the interferometer's output, or by entering and hence polluting the signal of any of the many photodiodes used to control the interferometer. Furthermore, light scattered off the main beam also affects the detector by directly inducing intensity fluctuations. Scattered light is also of major concern for the future, third-generation GW detectors: Einstein Telescope (ET)~\cite{Punturo_2010} in Europe, and Cosmic Explorer~\cite{evans2021} in the US.

Reducing SL noise is a difficult task both because it is a diffused phenomenon originating in, and potentially affecting, every part of the interferometer, and because it couples with mechanical noise in a non-linear way. A typical example is that of light scattered out of the main beam, reflected by a non-suspended, vibrating part, like the walls of a vacuum tank, and re-entering the main beam. In this process, the recombining SL acquires an additional phase that depends on the relative motion between the reflecting surface and the interferometer's optics, which can be considered at rest. The SL phase can then be expressed as the sum of a static term and a time-dependent one:
\begin{equation}
\phi_{SL} = \phi_0+ \phi_n (t) = \frac{4\pi}{\lambda} \left(x_0 +  x_n(t) \right)
\end{equation}
where $\phi_0$ is the static term and $\phi_n(t)$ is the time dependent-one. $x_0$ is the average path-length difference and $x_n (t)$ is the motion of the reflecting surface. Because the interference term between this light and the main beam depend sinusoidally on the phase, large $x_n(t)$ will appear distorted in a non linear way when converted into equivalent strain noise at the detector input \cite{Ottaway2012}. As a result, the SL noise appears as noise ``shoulders'' in the power spectral density (PSD) of the total strain noise of the detector and as arches in its time-frequency spectrograms \cite{LIGO:2020zwl}. Additionally, given the non-stationarity of the mechanical noise, SL noise is also non-stationary and can originate short-duration glitches of noise. 

Within a GW interferometer, SL can originate from a variety of mechanisms: light can be scattered by air molecules (this contribution is made negligible by having light propagate under high vacuum), by imperfections in the optical elements (residual surface roughness, bulk inhomogeneities), by contaminants deposited on the optical surfaces, by unstopped ghost beams (that is, beams originated by residual transmittivity of high reflectivity components or, vice-versa, by residual reflectivity of highly transparent ones), or can be produced in the clipping of laser beams.

\begin{figure}[h]
    \centering
    \includegraphics[width=0.95\linewidth]{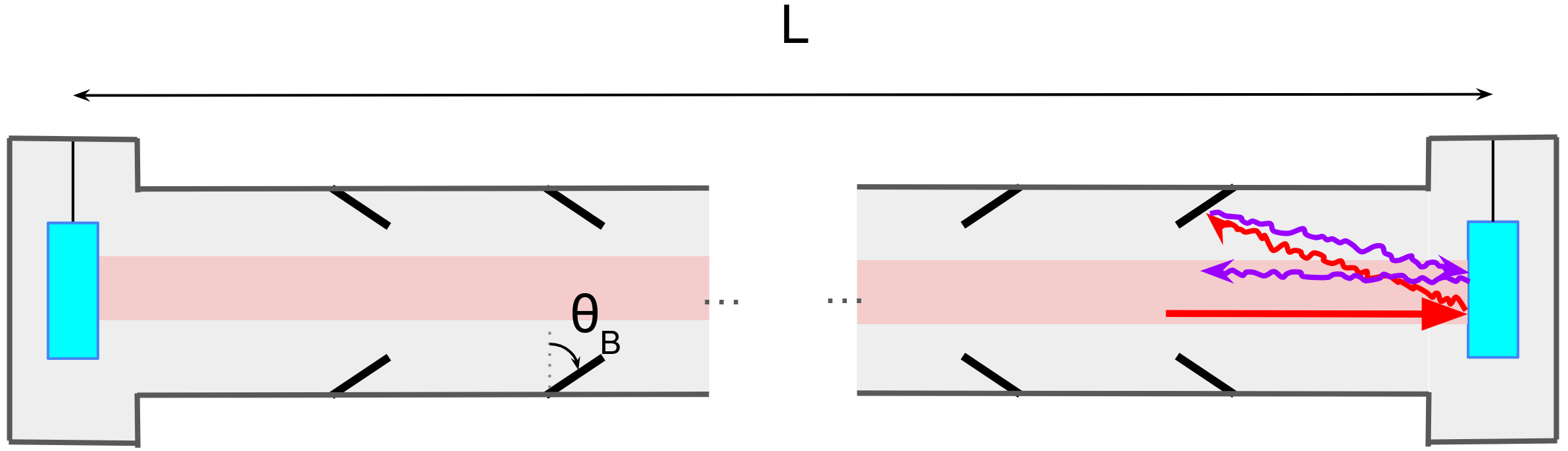}
    \caption{An arm of the ET interferometers consists of a Fabry-Perot cavity of length $L = 10$km. The input and output mirrors of such a cavity (the Test Masses, TM, pictured as light-blue rectangles) are suspended at the arm extremes in dedicated chambers connected by a vacuum pipe (1m diameter): concentric conical baffles (black thick lines) are distributed along the pipe. The baffle's inclination angle is $\theta_B = 55^\circ$. Dimensions from Ref.\cite{PhysRevD.108.102001}. In the right-most part of the scheme it is pictorially shown a light ray arriving at the TM (straight red arrow), being scattered towards a baffle (wiggled red arrow), scattered back from the baffle to the TM itself and finally re-entering the beam (wiggled purple arrows)}
    \label{fig:ETarms}
\end{figure}

The residual roughness of the interferometer's mirrors (the Test Masses, TMs) is an example of SL source due to surface imperfections. 
Polishing the Test Mass to a state-of-the-art level minimizes the generation and recoupling of SL: for instance, the residual roughness of the mirrors employed by the Virgo interferometric GW detector in O3 and O4 is 0.9 nm within the central area \cite{Degallaix:19} and even smoother surfaces are expected for the next observing run O5 and for ET. Nevertheless, the residual roughness still generates a non-negligible amount of SL that can recouple with the main beam and degrade performances if unstopped. To minimize recoupling, all present interferometric GW detectors have conical absorbing baffles distributed inside the arms following a suggestion dated back to 1989~\cite{Thorne1989}: their role is to completely shield the vacuum pipe surface, which has poor optical quality and comparatively large motion, from light scattered by the TMs at small angles.
Similarly, next-generation GW interferometers are planning the addition of baffles inside the arm pipes; see, for instance, ref.\cite{PhysRevD.108.102001} and Fig.\ref{fig:ETarms} for the case study of ET. In the case of the ET geometry, the baffles intercept light that is scattered by the TMs at angles between $\sim 10^{-2}\text{rad}$ and $\sim 10^{-5}\text{rad}$ from the optical axis, corresponding to the first and last baffles in the vacuum pipe, respectively \cite{PhysRevD.108.102001}.

While baffles are necessary in the arm pipes to absorb SL generated by the TMs, it was soon recognized \cite{FlanaganThorne} that they do not completely solve the problem, as their absorption is not perfect. Besides clipping the main beam, baffles can also scatter light which continues to propagate in the pipe. While light can reach the baffle and then recombine with the main beam through many different paths, here we focus on the most direct one: light originating from a TM and reaching a baffle can be scattered back by the baffle, reach the TM where it was generated, and be scattered again by the TM into the main beam (as represented in Fig.\ref{fig:ETarms}). Combined with the comparatively large vibration noise experienced by the non-suspended vacuum pipe, eventually amplified by internal resonances of the pipe itself and of the baffle, SL noise due to baffle scattering can become a major offender. 

Scattered light noise from baffle roughness has already been studied \cite{PhysRevD.108.102001}. In this work, we focus on another mechanism: light scattered by particle contaminants (here referred to as dust) deposited on the baffle surface. We point out that while cleanliness is a recognized issue and GW interferometers are developed in clean environments, to our knowledge, no one has yet conducted a quantitative analysis of stray light noise caused by dust in these instruments. This work aims to help fill this gap. 

As a case study, we focus on the arms of ET: this comes timely with the status of the project because cleanliness requirements for the fabrication and installation of the ET arm pipes are an important piece of information that helps refine the ET proposal. However, we note that cleanliness is an issue not limited to arm pipes alone, but is relevant for the entire instrumental facility. Hence, our work provides a way to address this problem in other cases as well.

We consider particles with diameters between $0.1 \,\mu$m and $300\, \mu$m. Larger particles should be rare in any clean environment, and their presence should be considered an outlier and treated separately.
Particles smaller than this range could, in principle, be quite numerous, but are difficult to quantify, as standard methods such as optical and digital microscopy fail; furthermore, clean rooms are not rated for particles smaller than $0.1 \mu$m~\cite{ISO14644}. Fortunately, they scatter little at the wavelengths of the ET lasers (1064 nm for the room temperature interferometers of Einstein Telescope, called ET-HF, and 1550 nm or about 2000 nm for the cryogenic temperature interferometers of Einstein Telescope, called ET-LF), and our estimations suggest that for plausible particle size distributions their contribution can be neglected compared to that of particles in the range indicated above.

This paper is organized as follows. In Section~\ref{sec:general} we discuss how dust deposited on a reflective surface scatters light, focusing on how this depends on light polarization, scattering angle, and particle size and refractive index. In Section~\ref{sec:ETcase} we specialize to the case of the baffles in the arm pipes of ET. 
In Section~\ref{sec:et_req} we set cleanliness requirements and suggest some guidelines for the installation of the ET arms and baffles. In Section~\ref{sec:concl} we conclude with ideas for future studies.

\section{Light scattered by particles on a surface}
\label{sec:general}
In this work, we compute scattering by particulates applying Mie theory, which uses exact solutions of Maxwell's equations in the case of monochromatic light scattered by spherical particles \cite{BohrenHuffman}. This theory is valid for all particle sizes, being particularly powerful in the case of particles comparable to the wavelength of light. In the limit of particles much smaller than the wavelength, Mie theory tends to Rayleigh scattering, whereas in the opposite limit of particles much larger than the wavelength, Mie theory leads to geometrical optics.

In a realistic environment, particles are not expected, in general, to be spherical, leading to different amounts and distributions of scattered light, as shown by both numerical \cite{Pierre_Drossart_1990} and experimental works \cite{Li:26, amt-7-931-2014}. However, for particles with a diameter comparable to or smaller than the wavelength of light, Mie theory is reported to be a good approximation even for moderately irregular particles; for larger particles, deviations from the spherical case produce results that are off by up to approximately one order of magnitude. In the case of this specific study, however, this variation is still relatively modest compared to that arising from lack of knowledge of the final particles' size and material distributions, which is several orders of magnitude, as will be shown later (see, for example, Fig.\ref{fig:brdf_gray_lines}, \ref{fig:brdf_vs_m_1}, and \ref{fig:brdf_vs_m_100}).

Given the impossibility of identifying, at present, a realistic particle shape distribution, and the modest impact of this choice on the overall uncertainty of the result, any correction to the model appears arbitrary and could lead to further departure from a realistic case. We thus choose to adopt the spherical approximation and use Mie theory.

With this assumption in mind, the general framework established here to evaluate cleanliness requirements can and will need to be refined once more information becomes available not only on the expected shape of the contaminating particles, but also on their size distribution and material composition.
%%\textcolor{gray}{The Mie theory is an exact solution of Maxwell's equations in the case of monochromatic light scattered by spherical particles  \cite{BohrenHuffman}. This theory is valid for all particle sizes, being particularly powerful in the case of particles comparable to the light wavelength. In the limit of particles much smaller than the wavelength, Mie theory tends to Rayleigh scattering, whereas in the opposite limit of particles much larger than the wavelength, Mie theory leads to geometrical optics.} 

Let us define for each scattering angle the scattering plane as the one containing the incident and scattered rays. The Mie theory allows to compute the relative amplitudes for the electric field components scattered by a particle:
\begin{equation}
\begin{pmatrix}
E^{s}_1 \\
E^{s}_2
\end{pmatrix} =
\begin{pmatrix}
S_{11} & S_{12} \\
S_{21} & S_{22}
\end{pmatrix}
\begin{pmatrix}
E^{i}_1 \\
E^{i}_2
\end{pmatrix},
\label{eq:mie_general}
\end{equation}
where $E^i$ and $E^s$ are the incident and scattered field in the proximity of the particle (i. e. we neglect here the field propagation term) and the subscripts $1 , 2$ refer to the polarizations perpendicular and parallel to the scattering plane, respectively. 
The Mie scattering matrix coefficients $S_{jk}\equiv S_{jk}(x, m, \theta)$, with $j, k=1, 2$, depend on the particle's reduced diameter $x=\pi \, D/\lambda$, where D is the particles diameter and $\lambda$ is the light wavelength; on the scattering angle $\theta$ with respect to the incoming beam; and on the complex index of refraction $m$ of the particle relative to the medium (which, in this work, we assume to be vacuum). It must be noted that the scattering coefficients $S_{jk}$ are complex as they affect both amplitude and phase of the incident field.

Let us consider a particle on a surface. For simplicity, here we consider only light scattered in the plane of incidence, which is adequate for our application (as discussed in Section \ref{sec:ETcase}); in this situation, the polarizations 1 and 2 become coincident, respectively, with the S and P polarizations defined for the surface. We also assume that particles are homogeneous and made of an isotropic material: hence, the anti-diagonal terms $S_{12}$ and $S_{21}$ describing polarization mixing are null in the plane of incidence \cite{spyak1}. Under this assumption, in the following we will drop one index for clarity, so that $S_1 \equiv S_{11}$ and $S_2 \equiv S_{22}$. We compute the scattering probabilities $S_{1}$ and $S_{2}$ with the python package \texttt{MiePython} \cite{miepython}.

\begin{figure}[!h]
    \centering
    \includegraphics[width=0.95\linewidth]{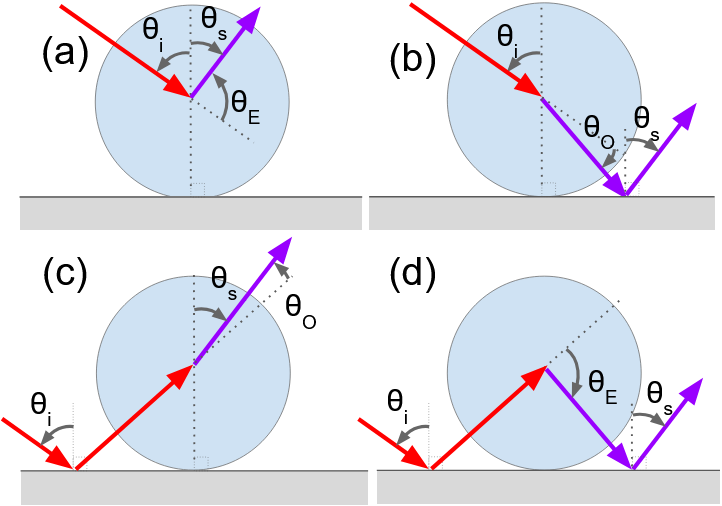} 
    \caption{Schematic representation of the four processes by which light reaching a particle deposited on a surface with an angle $\theta_i$ can be scattered at an angle $\theta_s$: (a) direct scattering; (b) scattering towards the surface and hence reflection; (c) reflection by the surface and then scattering; (d) reflection by the surface, scattering by the particle and reflection by the surface.
    Incident light is shown as thick red arrow, at an incident angle $\theta_i$ defined counterclockwise with respect to the surface normal. Scattered light is shown as purple arrows, eventually emerging at an angle $\theta_s$ defined clockwise with respect to the surface's normal. The scattering angle relevant for the particle interaction assumes two separate values $\theta_E$ and $\theta_O$ in the case of even (cases (a) and (d)) or odd (cases (b) and (c)) number of surface reflections, respectively.}
    \label{fig:3processes_h}
\end{figure}

Figure \ref{fig:3processes_h} shows how light can be scattered at an angle $\theta_s$ by a particle on a reflecting surface; we only consider the reflectivity of the surface and ignore its scattering (i.e., we neglect light that is scattered by the surface either before or after interacting with the particle). We outline that the processes (c) and (d) in Figure \ref{fig:3processes_h} have not always been considered in the literature (see for instance ref~\cite{spyak1, Fest}): however by the symmetry of the problem and of the scattering process these terms need to be included in the total scattering by particles on a surface (see Eq.\eqref{eq:brdf_dust_SS_coherent} and following).
Furthermore, we consider light interacting only once with the particles and we consider the four processes in Figure \ref{fig:3processes_h} as independent: in doing so, we neglect the fact that part of the light involved in one process is no longer available to participate in the others, leading us to overestimate the total effect by effectively counting some rays more than once, as they are considered to contribute to multiple processes in Fig.~\ref{fig:3processes_h}.

Light scattered in the plane of incidence by a surface is described by its bidirectional reflectance distribution function ($BRDF$):
\begin{equation}
\frac{d\mathcal{P}}{d\Omega} = \frac{1}{P_i} \, \frac{dP_s}{d\Omega} \, \frac{1}{\cos \theta_s}= BRDF
\label{eq:brdf_def}
\end{equation}
where $P_i$ is the incident power, $dP_s$ is the infinitesimal power scattered in the direction of the scattering angle $\theta_s$, defined with respect to the normal to the surface, in an infinitesimal solid angle $d\Omega$. In general the $BRDF$ is a function of the angle of incidence $\theta_i$ and $\theta_s$ - we neglect here any dependence on the azimuth angle as we consider isotropic surfaces. We indicate the bidirectional reflectance distribution function of a baffle as $BRDF_{baffle}$.

Let us now consider well-separated particles randomly distributed on a surface hit by light incident at an angle $\theta_i$ with respect to the surface normal: since their positions are random and unknown, we consider the total scattered light as the incoherent sum of all the single particle contributions. Let $f(x)$ be the density of the particles (i.e. the number of particles with reduced diameter $x$ per unit area), all having the same refractive index $m$; let $R_{S,P}(\theta_{s,i})$ be the power reflectivity of the surface for S- or P-polarized light and at the $\theta_s$ or $\theta_i$ angles of incidence of the light on the surface.

%In order to compute the $BRDF$ of the particles, $BRDF_{dust}$, for polarized light by coherently summing all the contributions, we need to account for two phase differences among them. Firstly, each time light is reflected by the surface it gains an additional $\pi$ phase. Secondly, in the cases presenting reflections (cases $b,\ c,\ d$) there is also a phase difference coming from the optical phase difference of the light rays being reflected with respect to the light rays interacting only with the particle. The path length difference can be computed from geometry (details in Appendix \ref{sec:brdf_coherent}).

For each single particle, the contribution of the four scattering paths depicted in Fig.\ref{fig:3processes_h} should in principle be added coherently. Appendix \ref{sec:brdf_coherent} shows the details of the calculation. However, the relative phase of the different scattering channels is governed by the distance between the center of the particle and the effective reflective surface. This is not well defined in realistic scenarios for at least two reasons: (i) real particles are not spherical, and the distance between their ``center'' (however defined) and the surface on which they rest can vary substantially, on the scale of the light wavelength, even for modest deviations from the spherical shape; (ii) for many surfaces, especially coated ones, the effective reflective plane is different from the physical surface on which the particle rests. These effects make the phase relation between the four terms unpredictable and particle-dependent, so that coherent effects average out for a distribution of realistic particles; we thus choose to adopt an incoherent sum.

Eventually, the $BRDF_{dust}$ due to dust for the S and P light polarizations are:
\begin{equation}
\begin{split}
BRDF&_{dust}^{SS}(\lambda, m, D, \theta_i, \theta_s) = \frac{\lambda^2}{4\pi^2 \cos(\theta_s)} \int f(x) ... \\ 
&\times \{ \left[R_S(\theta_s) + R_S(\theta_i)\right] \cdot |S_1(x, m, \theta_O)|^2 ... \\
& + \left[1 + R_S(\theta_s)R_S(\theta_i)\right] \cdot |S_1(x,m,\theta_E)|^2 \} \times dx
\label{eq:brdf_dust_SS_incoherent}
\end{split}
\end{equation}
\begin{equation}
\begin{split}
BRDF&_{dust}^{PP}(\lambda, m, D, \theta_i, \theta_s) = \frac{\lambda^2}{4\pi^2 \cos(\theta_s)} \int f(x) ... \\ 
&\times \{ \left[R_P(\theta_s) + R_P(\theta_i)\right] \cdot |S_2(x, m, \theta_O)|^2 ... \\
& + \left[1 + R_P(\theta_s)R_P(\theta_i)\right]\cdot  |S_2(x,m,\theta_E)|^2 \} \times dx
\label{eq:brdf_dust_PP_incoherent}
\end{split}
\end{equation}
where the apexes $PP$ and $SS$ indicate the polarization of the incident and scattered light. The angles $\theta_{E,O}$ define the direction relative to the incident light of light scattered by the particle, respectively, in case of an even ($k$ = E; cases $(a)$ and $(d)$ of Figure~\ref{fig:3processes_h}) or odd ($k$ = O; cases $(b)$ and $(c)$  of Figure~\ref{fig:3processes_h}) number of reflections by the surface. Geometry implies $\theta_E = \pi - \theta_s - \theta_i$ and $\theta_O = \theta_s - \theta_i$. For unpolarized incident and scattered light (apex UU), the $BRDF$ is obtained by averaging the expressions in Eq.\ref{eq:brdf_dust_SS_coherent}-\ref{eq:brdf_dust_PP_coherent}:
\begin{equation}
    BRDF_{dust}^{UU} = \frac{BRDF_{dust}^{PP} + BRDF_{dust}^{SS}}{2}
    \label{eq:brdf_dust}
\end{equation}

\subsection{Angular distribution}
\label{sec:gen_angle}
To illustrate the main features expected from the application of Eq.\eqref{eq:brdf_dust}, we start by studying two specific cases of scattering by small and large particles on a surface. For separating the different contributions, we introduce the symbol $BRDF_{dust}^{k,z}$ to distinguish between the $k = E, O$ parity  of the number of reflections and the $z=1,2$ polarizations. We also introduce the cosine-corrected $BRDF$, indicated as $cBRDF_{dust}^{k,z} \equiv BRDF_{dust}^{k,z} \times \cos{\theta_s}$. This quantity allows for a better visualization by avoiding the (mathematical, i.e. not physical) divergence of the scattering function for $\theta_s \rightarrow\pm \pi/2$ and originating from the cosine term in the definition of Eq.\eqref{eq:brdf_def}.

\begin{figure}[!h]
    \centering 
    \subfigure[$x=0.3$, $z = 1 $.]{
        \includegraphics[width=0.22\textwidth]{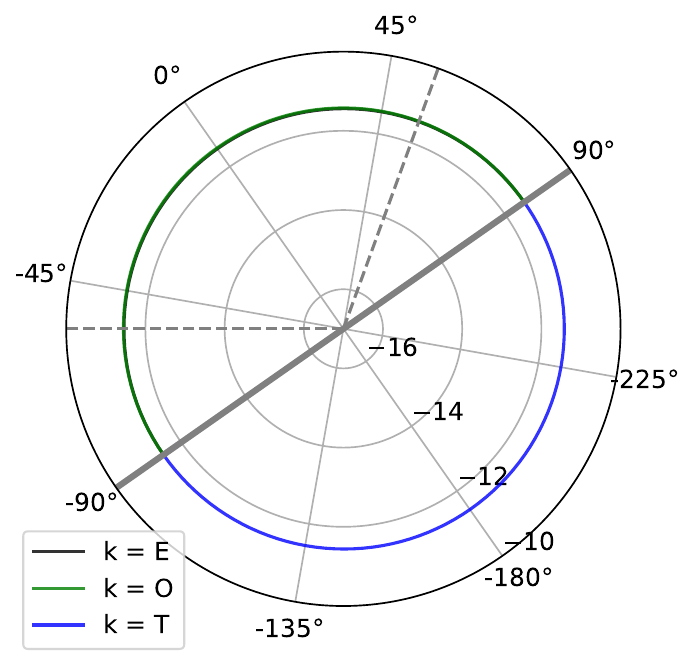} 
    }
    \hfill
    \subfigure[$x = 0.3$, $z = 2 $.]{
        \includegraphics[width=0.22\textwidth]{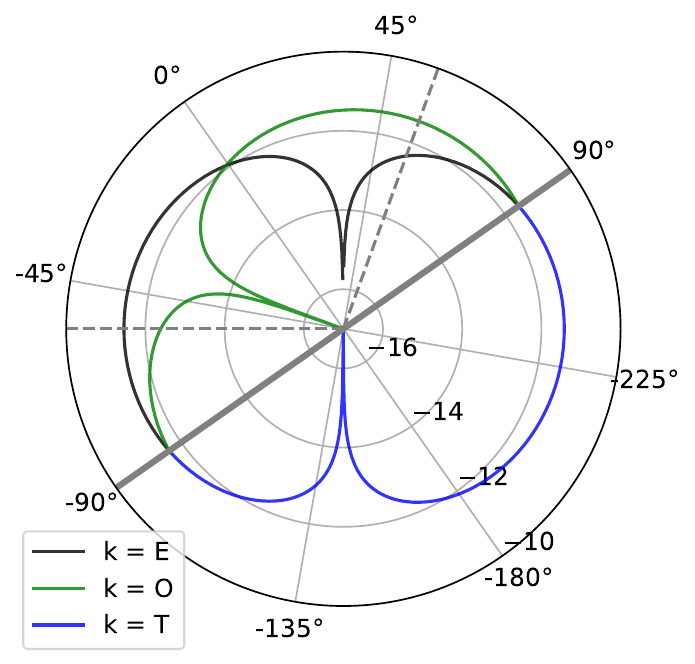} 
    }
    \par
    \subfigure[$x=300$, $z = 1 $.]{
        \includegraphics[width=0.22\textwidth]{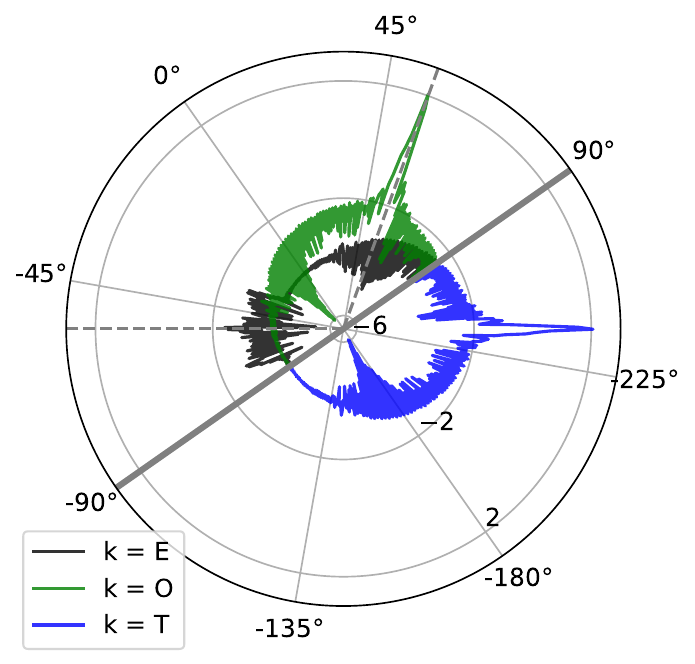}
    } 
    \hfill
    \subfigure[$x=300$, $z = 2 $.]{
        \includegraphics[width=0.22\textwidth]{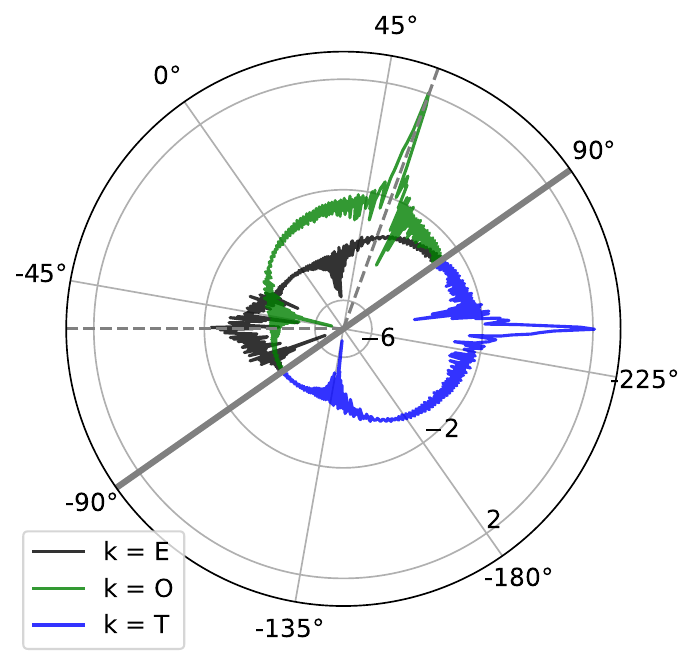}
    }
    \caption{Polar plot of $Log_{10} \left[cBRDF_{dust}^{(k, z)} \right] $ for $z=1$ (left plots) and $z = 2$ (right plots) for particles of reduced diameter $x=0.3$ (upper plots) and $x=300$ (lower plots). The dashed horizontal line is the direction of incident light; specular reflection direction is also shown as dashed. The black and blue traces together represent the scattering field from a suspended particle. However, if the particle is deposited on a reflecting surface, drawn as a diagonal thick gray line, the blue part of the scattered field is reflected and becomes the green trace. Black and green traces thus represent the $k=E$ and $k=O$ scattering contributions, respectively, and the blue trace is the scattered light as if it were fully transmitted by a transparent surface ($k=T$). All panels have $m=m^*\equiv1.5 + i \, 10^{-3}$ (see Sec.\ref{sec:ref_index}) and $R=1$, assumed to be constant across all angles and equal for both polarizations. The reflecting surface is tilted to represent an ET arm-pipe baffle, whose normal forms an angle $\theta_B = 55^\circ$ with respect to the incident light (horizontal). Angular axis spans the scattering angle $\theta_s$ whose origin is at the direction perpendicular to the surface. Due to the large difference in amplitudes, the plots for $x=3$ (a, b) and $x=300$ (c, d) use different radial ranges, but each pair (a, b) and (c, d) shares the same range.}
    \label{fig:brdf_polar}
\end{figure}

Figure \ref{fig:brdf_polar} illustrates, separately for small and large particles and for the two different polarizations, how the presence of a reflecting surface mixes the $k = E$ and $k = O$ scattering contributions; this helps to understand some of the features visible in the total scattering field.

The scattering from particles smaller than the light wavelength is approximated by Rayleigh scattering. (Fig.\ref{fig:brdf_polar} panels $(a),(b)$). For $z=1$ the scattering is broadly isotropic; for $z=2$, both the $O$ and $E$ contributions exhibits a direction in which they vanish, though not simultaneously, as the vanishing directions differ between the two cases.

Fig.\ref{fig:brdf_polar} panels $(c), (d)$ show the corresponding plots for a large particle (reduced diameter $x=300$), and $z=1, 2$. 
Two main features can be appreciated in this case: a peak at $\theta_s=\theta_i$ in the forward direction and a smaller one at $\theta_s=-\theta_i$ in the backward direction. The latter corresponds to retro-scattering by the particles (panel $(a)$ and $(d)$ in Fig.~\ref{fig:3processes_h}); the former corresponds to light scattered ahead by the particle and reflected by the surface, or vice-versa (these are the processes depicted in panels (b) and (c) in Fig.~\ref{fig:3processes_h}). This becomes evident by comparing the forward term with that expected if the surface were perfectly transparent, also shown in the figure. For ET, we are specifically interested in $\theta_s=-\theta_i$: Section~\ref{sec:ETcase} discusses this case in more detail. 

\subsection{Role of the refractive index}\label{sec:ref_index}
Particle scattering depends on the refractive index $m$, which in turn depends on the material and thus, in general, on the origin of the particles. Concerning ET, estimation of the most likely particle materials is not available. Moreover, it is still undecided where the detector will be located. Presently three possibilities are being considered: in the Sardinia island in Italy; at the border between Belgium, the Netherlands and Germany; and in Saxony. We expect that contaminants of environmental origin differ in the three cases and show different proportions of Saharan, ocean, urban, and land contributions. Furthermore, we need to consider that contaminants can also come from the tunnel excavation and construction, and from the fabrication, cleaning and handling of the ET components. 

Without any better knowledge of particle materials and to be conservative, the approach of the present work is to analyse a broad range of refractive indices: this provides a general framework that remains applicable in the absence of detailed information on the contaminant composition and can be readily specialized in future studies once the expected particle species are better characterized. In the following we allow the refractive index to vary within the broadest range that we found in the literature for environmental dust~\cite{Shettle79}. We also consider the case of the specific value $m^*=1.5 + i \, 10^{-3}$, which is often reported to be typical of environmental dust ~\cite{Fest, spyak4}: for instance, for $\lambda \sim 1 \, \mu$m, airborne mineral dust is reported to have $\mathrm{Re}(m)$ between 1.45 and 1.55 and $\mathrm{Im}(m)$ between $10^{-4}$ and $10^{-2}$~\cite{DiBiagio_dust}. 

Additionally, on the basis of the elemental analysis performed on dust found in the Virgo and LIGO \cite{10.1117/12.2047327} interferometric GW detector, we report results also for human skin and aluminum: aluminum is among the most common dust materials found in Virgo; copper is also found commonly but its refractive index is similar to that of aluminum; their refractive indexes are not contained in the range of environmental dust. In Virgo, organic materials are also commonly found: we represent them collectively as human skin. Table~\ref{tab:m_values} summarizes all refractive indexes we consider in this work.

\begin{table}[]
    \centering
\resizebox{\linewidth}{!}{\begin{tabular}{lcc}
    \hline
  & Re($m$)& Im($m$) \\
    \hline

Environmental dust & [1.3, 1.8] & [$10^{-6}$, 0.8]  \\
$m^*$ & 1.5 & 0.001\\
Human skin & 1.42 & 0.007\\
Aluminum ($\lambda$ = 1064 nm) & 1.25 & 10.5 \\
Aluminum ($\lambda$ = 1550 nm) & 1.47 & 16.1 \\
Aluminum ($\lambda$ = 2000 nm) &  2.20 & 21.0 \\
\hline
\end{tabular}}
\caption{Real and imaginary parts of the complex refractive index $m$ used in this work. For environmental dust (second row), we consider both components to vary within the maximum range reported in~\cite{Shettle79}; the ``typical'' value $m^*$  is given in the third row and that of human skin in the fourth row~\cite{skin}. For simplicity, the same value is assumed at 1064~nm, 1550~nm and 2000~nm in all these cases. The complex refractive index of aluminum~\cite{Ordal:88}, which varies sensibly with wavelength, is listed in the last three rows, specified separately for each ET wavelength.}\label{tab:m_values}
\end{table}
To highlight the dependence of the scattering on particle size and material, we consider particles having specific reduced diameters ($x = 0.3, 1, 3, 10, 30, 300$) and present with the same numerical density $f(x)$, namely 1 particle over 1~mm$^2$; our results are easily scalable to other densities given the linearity of Eq.~\eqref{eq:brdf_dust} with $f(x)$. 

For environmental dust we divide the refractive index space $\mathrm{Re}(m) \times \mathrm{Im}(m)$ reported in Table~\ref{tab:m_values}, in a $26 \times 19$ grid. In detail, $\mathrm{Re}(m)$ is evenly spaced between (1.3, 1.8) with a step size of 0.2; $\mathrm{Im}(m)$ is instead sampled with logarithmic spacing in order to cover its many orders of magnitude, and assumes the values 
$[1, 2.5, 5] \cdot 10^{-g}$, with $g$= 1, .., 6  and the value 0.8. 

\begin{figure*}[!htbp]
    \centering
    \subfigure[]{
        \includegraphics[width=0.45\textwidth]{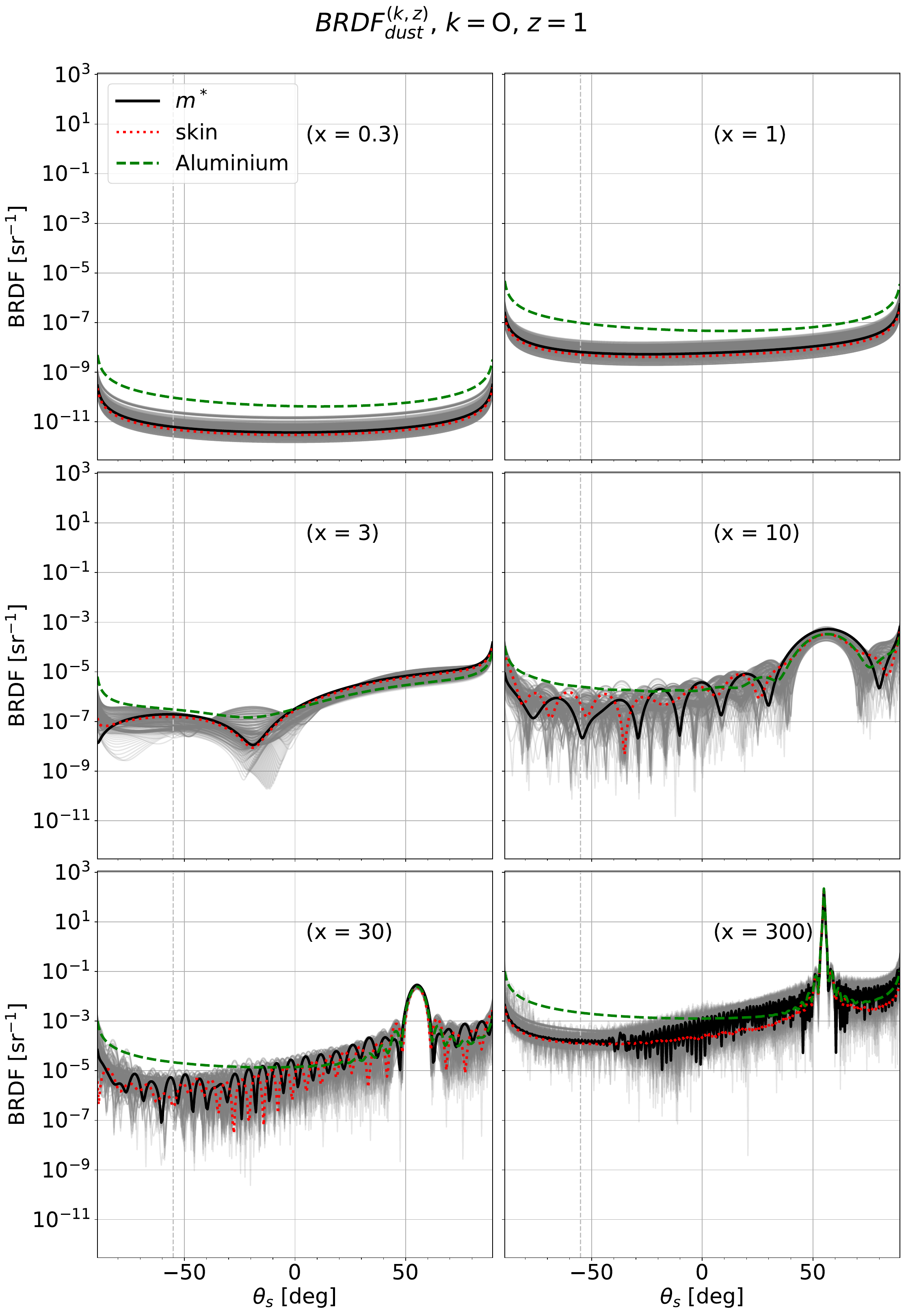}
        \label{fig:brdf_vs_th_f1}
    }
    \subfigure[]{
        \includegraphics[width=0.45\textwidth]{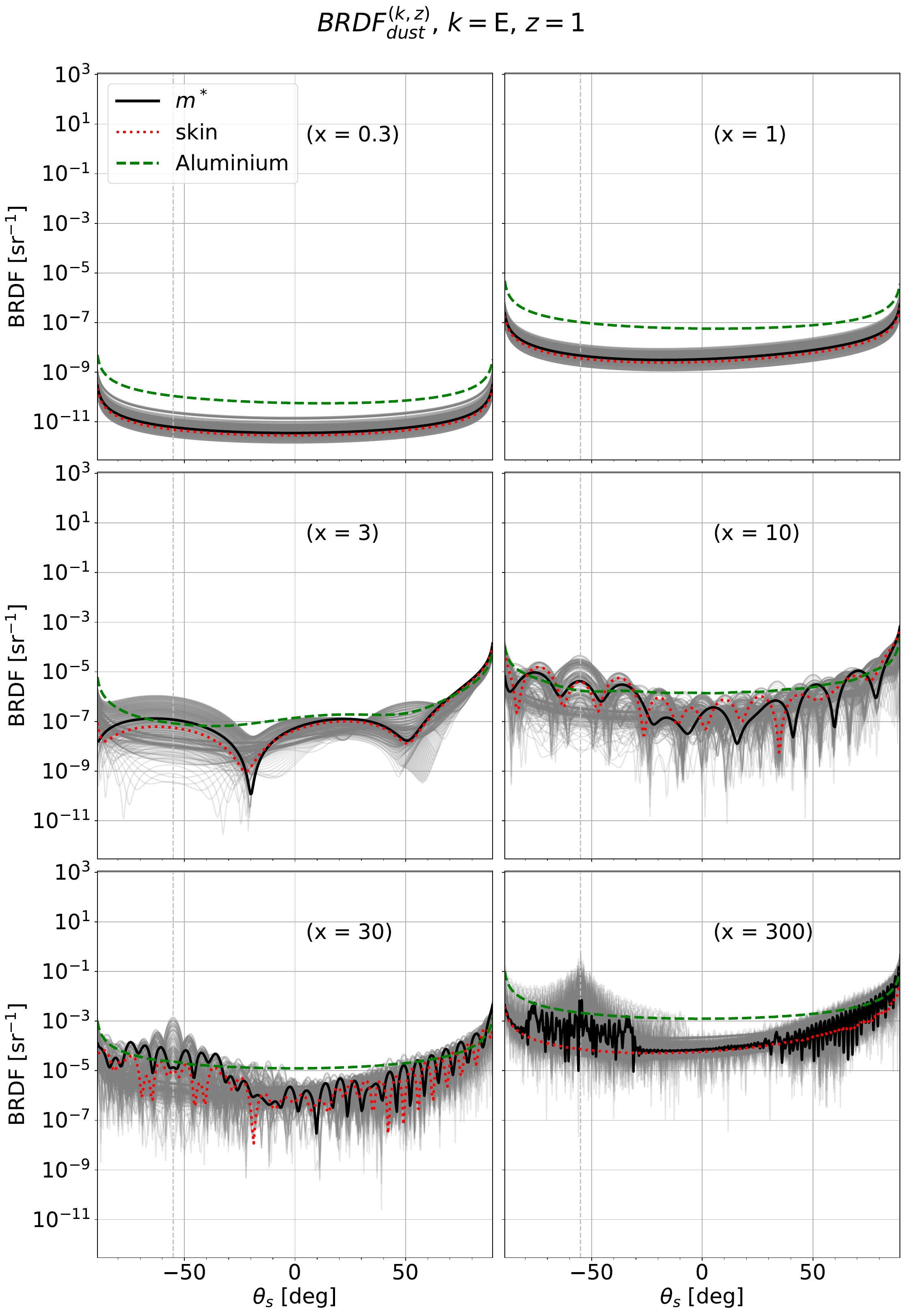}
        \label{fig:brdf_vs_th_b1}
    }
    
    \subfigure[]{
        \includegraphics[width=0.45\textwidth]{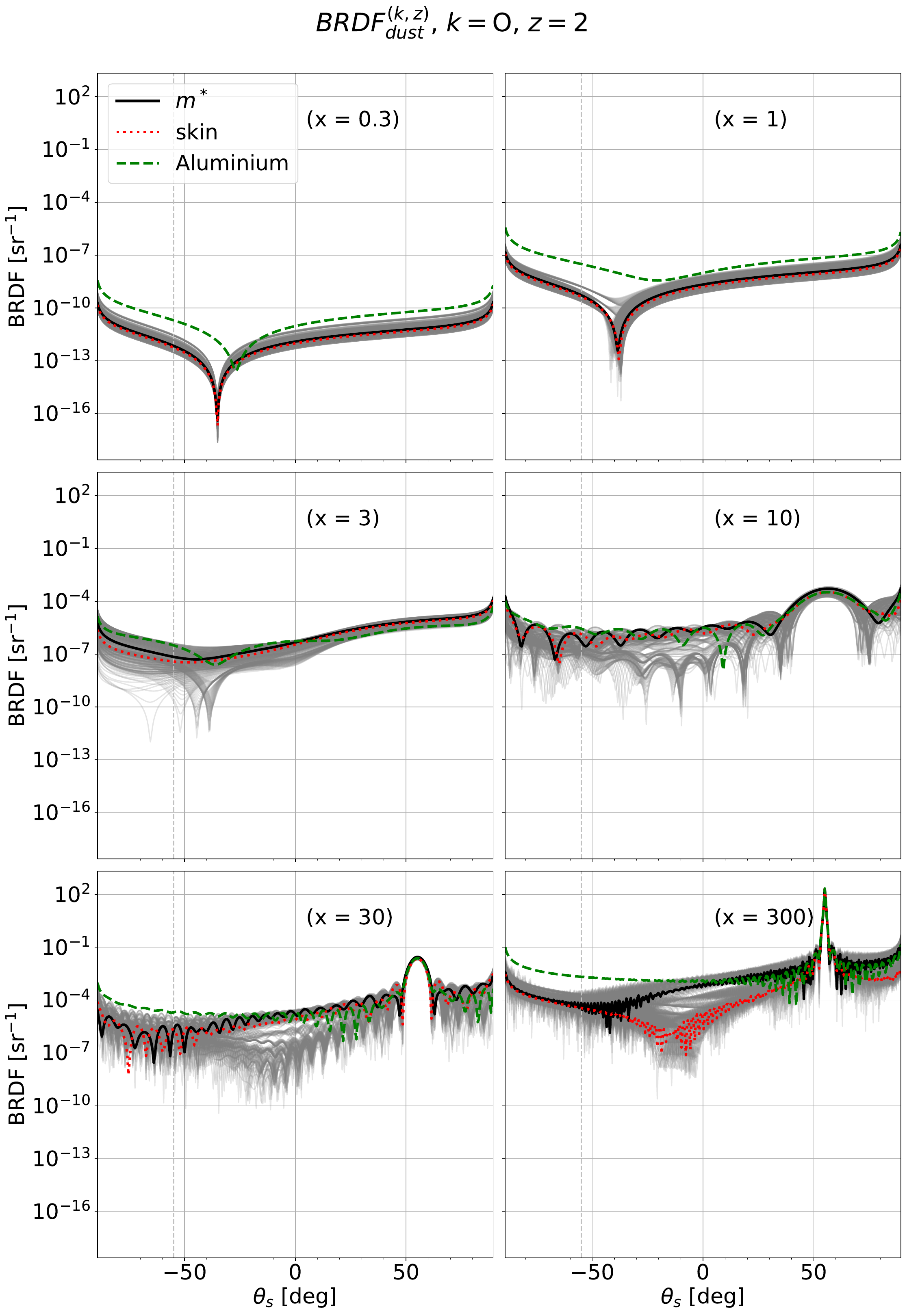}
        \label{fig:brdf_vs_th_f2}
    }
    \subfigure[]{
        \includegraphics[width=0.45\textwidth]{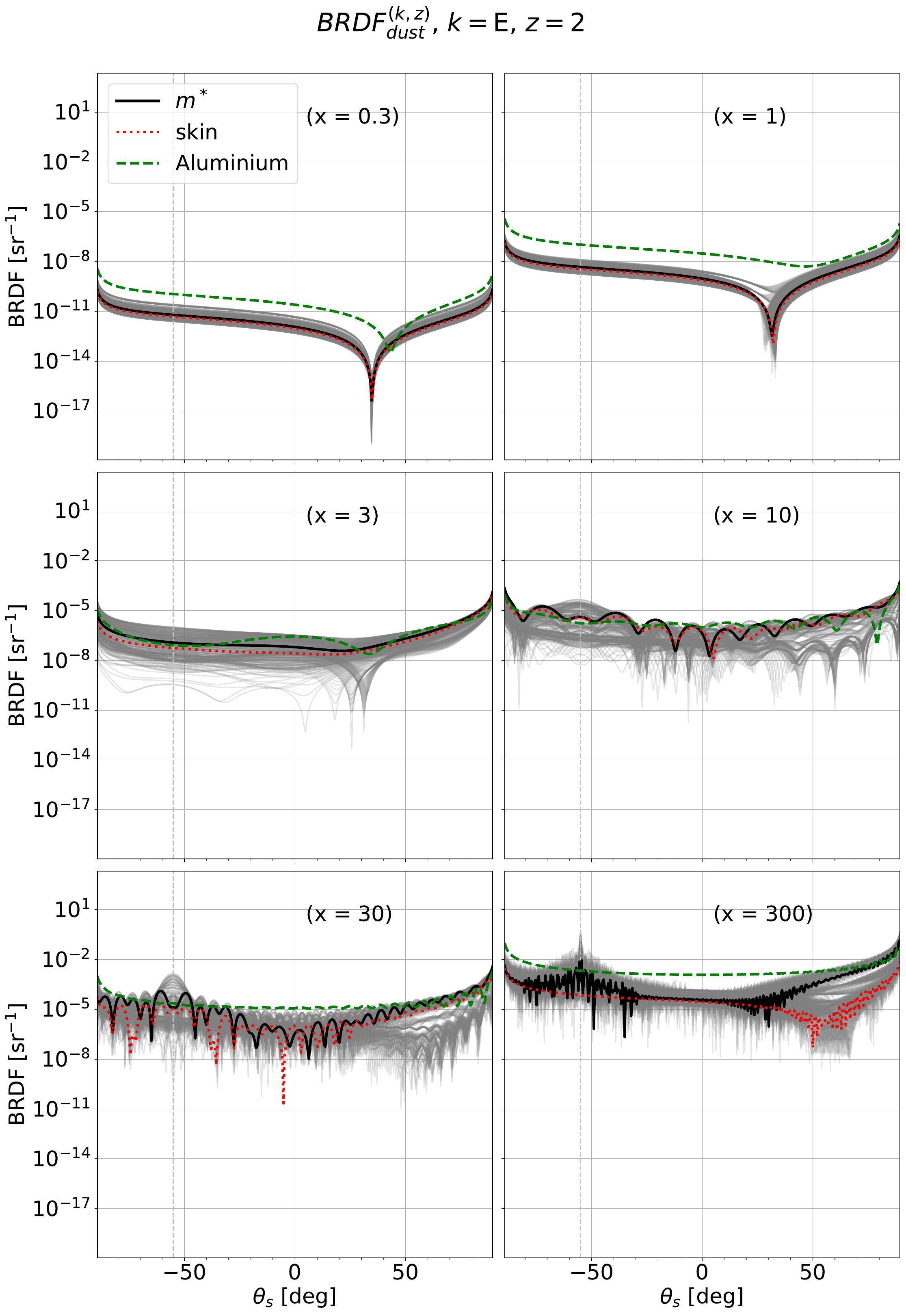}
        \label{fig:brdf_vs_th_b2}
    }
    
    \caption{BRDF contributions corresponding to odd ($k=O$) and even ($k=E$) number of reflections from the surface, for the two polarizations $z=1,2$, as a function of the scattering angle $\theta_s$. The angle of incidence is $\theta_i=\theta_B$ and the surface reflectivity is $R_{S,P}(\theta)=1$ over all angles and for both polarizations. The BRDF is computed for $\lambda=1064nm$. The vertical gray dashed line marks $\theta_s=-\theta_i$. The curves corresponding to all the refractive index values in the $26 \times 19$ grid are shown in gray for particles of reduced diameters $x$ ranging from 0.3 (top left) to 300 (bottom right), as indicated in each panel. The curves corresponding to the refractive indexes $m^*$ (black solid line), of the human skin (red dotted line), and of aluminum (green dashed line) are also shown. In all cases the particles density is $f(x)=10^{6}$ particles/$m^2$.}
    \label{fig:brdf_gray_lines}
\end{figure*}

For each particle size $x$, using Eq.~\eqref{eq:brdf_dust} we compute the $BRDF_{dust}^{(k,z)}(\theta_i=\theta_B, \theta_s)$ at all refractive indexes in the grid, separately for the $k= E,O$ cases and the two polarizations $z=1, 2$. For each $x$ we obtain 4 sets of curves, shown in Figures~\ref{fig:brdf_vs_th_f1}-\ref{fig:brdf_vs_th_b2}. In these plots we consider the underlying surface to be perfectly reflective, i.e. $R=R_S(\theta)=R_P(\theta)=1$ in Eq.~\eqref{eq:brdf_dust_SS_incoherent}-\eqref{eq:brdf_dust_PP_incoherent} for any angle $\theta$. While this assumption is not realistic, it allows to compare more easily the odd and even contributions. Figures \ref{fig:brdf_vs_th_f1}-\ref{fig:brdf_vs_th_b2} show also the curves computed for $m=m^*$, human skin and aluminum.

A common feature is that a series of ripples appears with increasing diameters: these ripples originate from interference effects, at each individual particle, between different scattered rays (see also Figure~\ref{fig:brdf_polar}); because of its high $\mathrm{Im}(m)$, which quickly extinguishes multiple reflections/refractions inside the particle, aluminum does not show such a pattern. The angular position and separation of the local minima and maxima depend on the particle diameter and on the refractive index; thus, these ripples quickly average out in the case of an ensemble of particles with continuous distributions in size and index of refraction.

As already observed in sec.\ref{sec:gen_angle}, for small ($x \lesssim 1$) particles and  polarization $z=1$ (top panels in Fig.~\ref{fig:brdf_vs_th_f1}-\ref{fig:brdf_vs_th_b1}), the scattering function is isotropic and assumes comparably low values, increasing with dimension: small particles (with respect to light wavelength) scatter little and with no preferred direction (see also Figure~\ref{fig:brdf_polar}a). For $z=2$ (top panels in Fig.~\ref{fig:brdf_vs_th_f2}-\ref{fig:brdf_vs_th_b2}), the scattering pattern shows a deep minimum, as (already seen in Figure~\ref{fig:brdf_polar}b) whose angular position depends on both the particle's diameter and refractive index.
The behavior for large particles is instead more complicated, due to the presence of many interference peaks that strongly depend on both size and index of refraction.

\section{Scattering by dust on the baffles of the ET arm pipes}\label{sec:ETcase}
In ET, as in all GW observatories, the entire instrument, including the several-kilometer-long arms, is kept under high or ultra-high vacuum to minimize noise due to pressure fluctuations and other effects, as well as to avoid contaminating the mirrors. However, particles can enter the system at different moments before the instrument is evacuated, as, for example, during manufacturing, assembly, or even pumping and venting operations. Dust can circulate even after vacuum is established: vacuum system operations, mechanical shocks on tube walls, and operating gate valves can release particles that deposit onto various surfaces. Here we estimate the stray light noise caused by dust on the arm baffles, and set acceptable limits. We do not discuss the effect of particles on the TM surface.

The dominant process we consider is that of a photon reaching a baffle from a TM (either because it has been scattered or because it belongs to the weak tail of the Gaussian beam that is clipped by the baffle aperture), and being scattered by a dust particle directly towards the same TM, which scatters it again into the main beam. Any other path from a TM to a baffle and then to a TM again (and eventually the main beam) implies additional scattering events by either the vacuum pipe walls and/or the baffles and it is thus strongly suppressed. We do not discuss processes involving baffle edges.

Scattered rays reaching a TM from any of the events depicted in Fig.\ref{fig:3processes_h} undergo the same processes as those scattered by a clean baffle. Thus, to compute the total strain noise from contaminated baffles back-scattering towards the TMs, we can still use Eq.\ref{eq:brdf_def} replacing:
\begin{equation}
%\begin{split}
BRDF_{baffle}   \rightarrow BRDF_{baffle} +BRDF_{dust} \label{eq:BRDFbaffle}
%\end{split}
\end{equation}
When applying Eq.\ref{eq:BRDFbaffle} the dust is assumed to cover a relatively small percentage of the surface, so that the amount of light scattered directly by the surface without interacting with dust particles can be considered unchanged.

Given the small dimension of the TM (order 0.5m in diameter) compared to its distance to any of the baffles (from 100m to several km), the scattering angle from a point on a baffle to any point on the TM varies by less than $0.6^{\circ}$. Hence, the light that reaches the TM is essentially backscattered in the same direction it comes from: $\theta_s \sim -\theta_i$, which implies $\theta_E \sim -\theta_i$ and $\theta_O \sim \pi+\theta_i$ for the four cases of Fig.\ref{fig:3processes_h}. Under this geometrical assumption, the scattering and incident plane can be taken as coincident, as assumed in Sec.\ref{sec:general}.

For the sake of definiteness, we arbitrarily set a requirement on SL noise by dust to be equal to that from a clean baffles, and find the corresponding maximum allowable particle densities:
%%\textcolor{gray}{To set a requirement on SL noise by dust, we take scattering from clean baffles as a reference, and find maximal particle densities that satisfy}
\begin{equation}
BRDF_{dust} (\theta_i, \theta_s) =BRDF_{baffle }(\theta_i, \theta_s)
\label{eq:condition}
\end{equation}
for $\theta_s=-\theta_i$. For ET: $BRDF_{baffle } (\theta_i, -\theta_i)=10^{-4}$/str and $\theta_i = \theta_B = 55 ^{\circ}$ \cite{PhysRevD.108.102001}. It is important to stress that defining a working requirement for light scattering from dust in the ET arm-pipes is outside of the scope of this paper, and the choice made in Eq.\eqref{eq:condition} is aimed at establishing a convenient reference value and to illustrate the methodology. Owing to the linear dependence of the scattered light on the particle density (Eq.\ref{eq:brdf_dust_SS_incoherent}-\ref{eq:brdf_dust_PP_incoherent}), the resulting cleanliness requirements can be easily scaled to any BRDF constrain.

Of course, there is not a unique particle density $f(D)$ that satisfies Eq.\ref{eq:condition}, as different materials, sizes and numerosity combinations can produce the same total scattering.
In this section we investigate how the $BRDF$ of contaminated baffles depends on the particle size and refractive index: in particular, we study how light incident at an angle $\theta_B$ on a plane surface is backscattered by a set of spherical particles deposited on the surface. We consider light polarized either perpendicularly ($z=1$) or parallel ($z=2$) to the scattering plane.

Equations \eqref{eq:brdf_dust_SS_incoherent} and \eqref{eq:brdf_dust_PP_incoherent} show how the $BRDF$ from a distribution of particles on a surface is also dependent on the reflectivity $R$ of the surface itself, a beampipe baffle in our case. 
While a small $R$ makes the contribution of process (d) of Fig.\ref{fig:3processes_h} essentially negligible, as it represents a correction of order $R^2$ to the contribution of process (a), the same is not necessarily true for processes (b) and (c): in fact, the difference in scattering amplitudes at the two angles $\theta_{O,E}$ can make these processes' contribution to the total scattering comparable to that of precess (a) even if they are multiplied by a relatively small reflectivity coefficient. Since $\theta_s=-\theta_i$ in our case, we have $R(\theta_i) = R(\theta_s) = R$ : for simplicity and ease of comparison between the different contributions, in this section we assume $R = 1$ and independent of the polarization.

\begin{figure}[!h]
\includegraphics[scale=0.22]{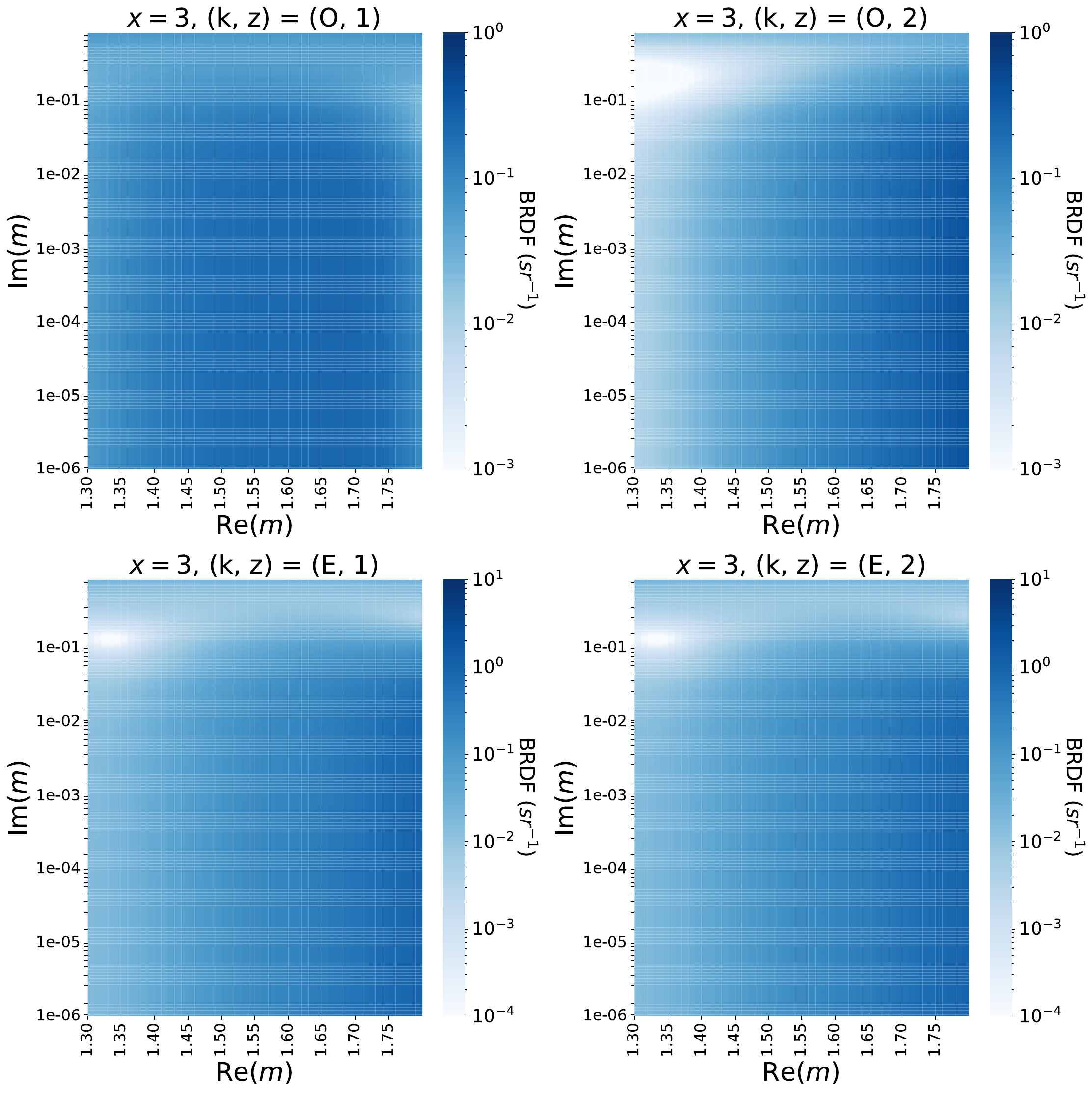}
\caption{Heatmaps of $BRDF_{dust}^{(k,z)}$ at $\theta_s=-\theta_i=-\theta_B$ against $\mathrm{Re}(m)$ in the horizontal axis and $\mathrm{Im}(m)$ in the vertical axis, for $(k, z)$ indicated in the plot titles. All plots assume $R=1$ and a distribution of particles with $x=3$ and density $f(x) = 10^{6}$~particles/m$^2$. Plots in the same row share the same color scale, displayed on the right.} 
\label{fig:brdf_vs_m_1}
\end{figure}

\begin{figure}[!h]
\includegraphics[scale=0.22]{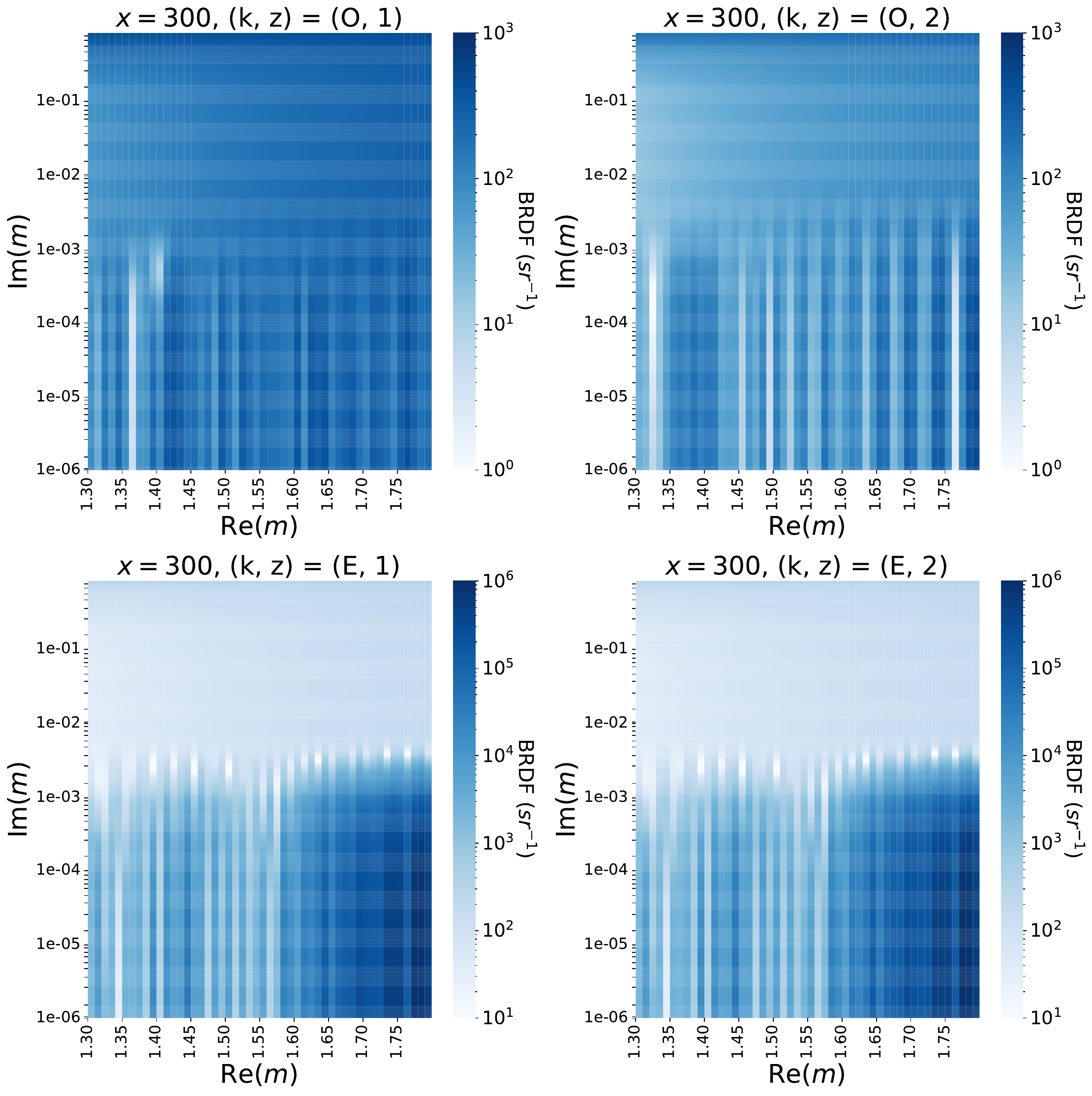}
\caption{Same as Figure~\ref{fig:brdf_vs_m_1} but for $x=300$.} \label{fig:brdf_vs_m_100}
\end{figure}

Building on the results of Sec.\ref{sec:ref_index} and focusing on the scattering geometry described above, here we highlight the range of (complex) refraction indices (from the values of Tab.\ref{tab:m_values}) that are more detrimental for ET: we focus on particles with $x=0.3$ (Fig.\ref{fig:brdf_vs_m_1}) and $x=300$ (Fig.\ref{fig:brdf_vs_m_100}) as examples of small and large particles respectively. We present the results separately for scattering reflection parities $k = O,E$ and polarizations $z = 1, 2$.

Again we see that small particles scatter less than large ones. Moreover, scattering is less dependent on the refractive index, especially for $(k,z)=(O,1)$ (see also Figure~\ref{fig:brdf_polar}a), and increases with higher $\mathrm{Re}(m)$ and lower $\mathrm{Im}(m)$ (Fig.~\ref{fig:brdf_vs_m_1}).

For large particles (see Fig.~\ref{fig:brdf_vs_m_100}) the behavior with $m$ is more complex: in addition to the general pattern of increasing scattering for higher values of $\mathrm{Re}(m)$ and lower values of $\mathrm{Im}(m)$ and reduced variability for $(k,z)=(O,1)$, we can see a series of ripples, with local minima and maxima, also observed in Sec.~\ref{sec:gen_angle}: they result from interference effects at the level of single particles, caused by specific combinations of $x$ and refractive index: the real part of $m$ affects the location of the extremes and the imaginary part affects their amplitude.

\begin{figure}[!h]
\includegraphics[width=1\columnwidth]{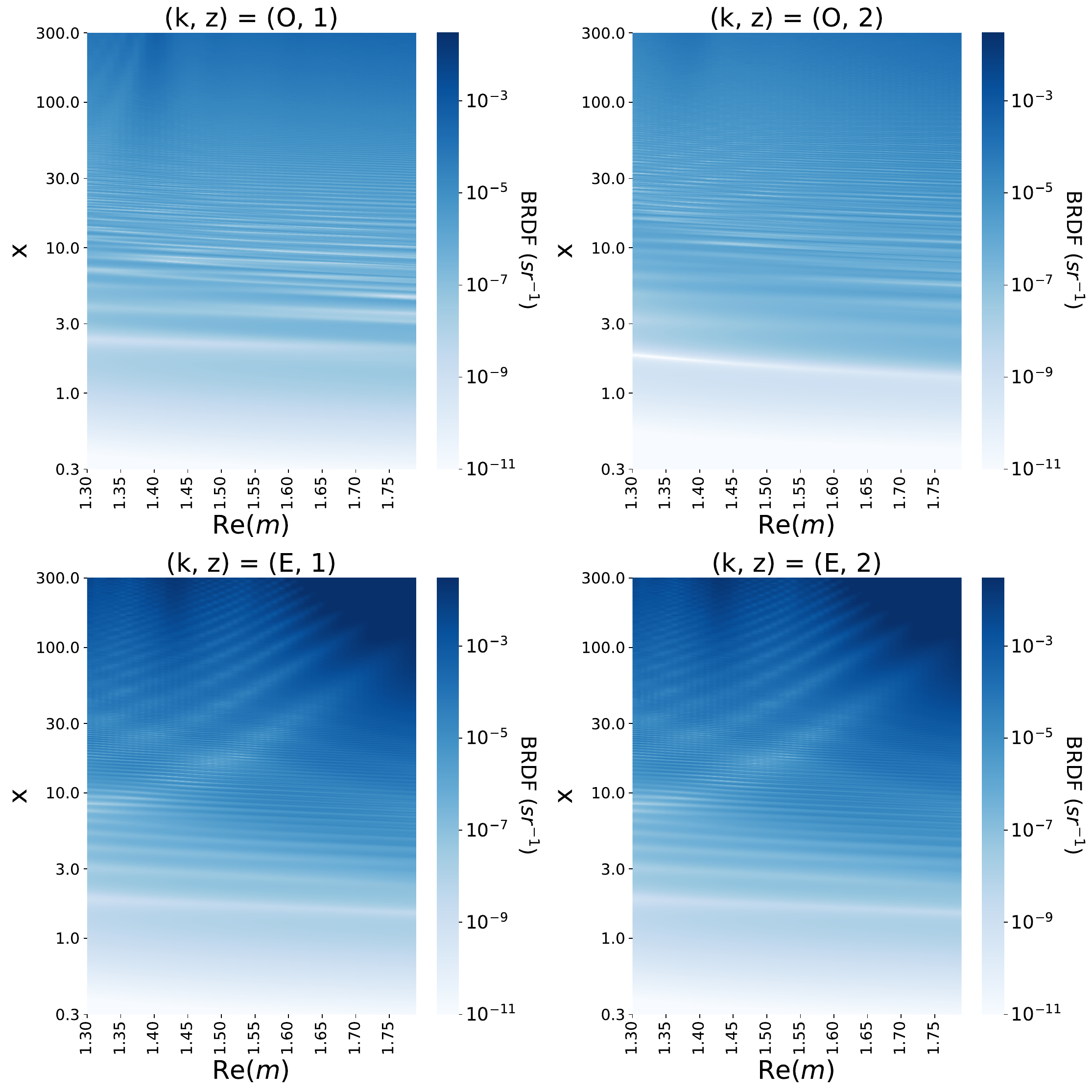}
\caption{Heatmaps of $BRDF_{dust}^{k,z}$ at $\theta_s= -\theta_i$, against $\mathrm{Re}(m)$ in the horizontal axis and $x$ in the vertical axis, for $(k, z)$ indicated in the plot titles. All plots refer to $\theta_i= \theta_B$,  $R=1$ and $\mathrm{Im}(m)=10^{-6}$. The same color code is used in all plots and is shown to their right.}
\label{fig:brdf_x_vs_rem}
\end{figure}

However, we note that we do not expect a realistic particle distribution to show such clear features. First of all, real particles are not in general perfectly spherical as assumed by Mie theory, and their scattering pattern would not exhibit such regular interference effects \cite{Huang2021}. In addition, as discussed in Sec.\ref{sec:et_req}, variations of just a few percent in the refractive index or particle size (the phase shift being proportional to $m \cdot x$) are enough to turn a local minimum into a local maximum, smoothing out these structures when considering an ensemble. Fig.\ref{fig:brdf_x_vs_rem} shows $BRDF_{dust}^{(k,z)}$ as a function of $\mathrm{Re}(m)$ and of the reduced diameter $x$. To highlight any ripple effects here we take $\mathrm{Im}(m)=10^{-6}$ (i.e. particles with the lowest absorption). Except for the smallest particles, varying $\mathrm{Re}(m)$ leads to an alternating pattern of maxima and minima, consistently with what is observed in Fig.\ref{fig:brdf_polar}, \ref{fig:brdf_gray_lines}, \ref{fig:brdf_vs_m_1} and \ref{fig:brdf_vs_m_100}.

This study indicates that, for ET, the most problematic particles are made of materials with high $\mathrm{Re}(m)$ and low $\mathrm{Im}(m)$. Glasses --- e.g. for BK7 $m=1.5066+i\cdot 1.0888 \cdot 10^{-8} $ at $\lambda=1064$  nm \cite{material_db} --- are examples: they are used widely in GW interferometers and laboratories and hence are likely to be found as dust there. Airborne mineral dust also has similar refractive index characteristics.

\section{Requirements and recommendations for ET}\label{sec:et_req}
We now refine the discussion by incorporating additional technical details needed to establish cleanliness requirements for the fabrication and installation of the arm pipes with conical baffles.

Our goal is to quantify the maximum number and size of particles that can be tolerated on baffle surfaces. This, in turn, will determine the required cleanliness class of, and allowable exposure times to, ET environments; it will also inform parts production protocols and work procedures.

We examine separately the different parameters that influence the scattering, as described in Eq.\eqref{eq:brdf_dust_SS_incoherent}-\eqref{eq:brdf_dust_PP_incoherent}, and thus affect the cleanliness requirements.  First, we consider the optical properties of the baffle material and the polarization of scattered light; these topics are addressed in Sec. \ref{sec:R_and_pol}. Next, we consider the properties of dust: as discussed in Sec.~\ref{sec:gen_angle}, Sec.~\ref{sec:ref_index} and Sec.\ref{sec:ETcase}, different combinations of particle size, refractive index, and numerosity can satisfy Eq.~\eqref{eq:condition}. A realistic scenario must account for a distribution of particles sizes and materials: in Sec.\ref{sec:dist}, we estimate a plausible particle size distribution and corresponding numerosity and we describe how we account for different refractive indices.
With these inputs, in Sec.\ref{sec:req}, we run our simulation to derive cleanliness requirements for the ET baffles under realistic conditions. In Sec.\ref{sec:guidelines} we discuss the practical implications of our findings for the assembly and installation of the ET beam pipes.

\subsection{Reflectivity and polarization}
\label{sec:R_and_pol}
The reflectivity $R$ for the ET baffles has not yet been defined. Previous studies on materials and coatings suggest that $R$ may lie in the range $0.5$ to $\sim 10^{-6}$ at $\theta_i \sim \theta_B$ \cite{LIGO_report_R}. However, given the large number of baffles ($\sim 100-300$) to be installed in each ET arm \cite{PhysRevD.108.102001}, the choice will likely be guided not only by optical performance, but also by cost, handling and procurement considerations. For the purpose of setting cleanliness requirements, we consider two representative cases: $R=1$ and $R=10^{-2}$. The first provides a conservative estimate, corresponding to the highest possible scattering, while the second is a more realistic assumption if the baffle reflectivity is effectively suppressed. Further reductions in $R$ do not substantially affect the results: the odd-reflection contribution ($k=O$) scales linearly with $R$ and thus vanishes for $R \to 0$, whereas the dominant even-reflection term ($k=E$) scales with $1+R^2$ and is only weakly dependent on $R$ in this regime.  

Ideally, one would need to know the exact position of each particle, project the incident polarization onto the particle’s scattering plane, and compute the corresponding Mie matrix coefficients. However, if the particles are homogeneously distributed over the conical baffle surface centered along the ET beam propagation axis, summing the contributions from all particles of a given size at random positions is equivalent to averaging over all possible orientations of the scattering plane — and therefore over polarization (see Appendix~\ref{sec:pol} for a mathematical justification).

In realistic conditions, the assumption of uniform distribution is likely too strong: indeed, for those deposition mechanism that are strongly influenced by gravity, the cylindrical symmetry is broken, leading to higher particle densities on the lower half of the baffles, and consequently introducing an asymmetry in the polarization of scattered light. On the other hand, the polarization properties of scattered light are strongly influenced by particle morphology \cite{Pierre_Drossart_1990}. In particular, deviations from the ideal spherical shape generally reduce the degree of polarization of the scattered field. Since real contamination is expected to consist of irregularly shaped particles, the polarization degree may be weakly maintained. The extent to which this effect compensates for the asymmetric particle deposition requires a dedicated quantitative investigation and is beyond the scope of the present study.

With this in mind, in the present work, cleanliness requirements are evaluated for the unpolarized case, using $BRDF^{UU}$ as defined in Eq.~\ref{eq:brdf_dust}. To simplify notation, the superscript $UU$ is omitted in the following. Finally, we incoherently sum the contributions from scattering with odd and even numbers of reflections $k = O, E$, as explained in Sec.~\ref{sec:ETcase}.

\subsection{Size-numerosity for exposure in cleanrooms}
\label{sec:dist}
Since we are interested in requirements for cleanrooms, we refer to the ISO-14644 standard \cite{ISO14644}, which quantifies the maximum allowed volumetric particles concentration for a cleanroom of ISO class $N$ via a cumulative distribution $C_N(D)$ following a power law, valid for particle diameters $D$ between 0.1 and 5 $\mu m$:
\begin{equation}
    C_N(D) \leq 10^N \times \left(\frac{0.1}{D}\right)^{2.08}
\end{equation}
We assume that the same power law is valid in the entire size range we consider, i.e. also for particles larger than $5\mu m$.

As a reference, we assign to the baffle the contamination expected on a flat surface of the same area, assuming horizontal exposure to the maximum allowed volumetric particle concentration in an ISO 7 cleanroom. To do so, we calculate the surface contamination by multiplying the volumetric particle concentration by the size-dependent particle deposition velocity from Ref.~\cite{WhyteAgricola} and by the exposure time $t$: this has been experimentally validated up to $25\mathrm{\mu m}$ (see Ref.\cite{moscatello_phd}, Sec.5.1).

We note, however, that large particles tend to settle fast, making their population strongly dependent on the specific cleanroom layout, nearby activities, handling procedures, personnel movements, and other process-dependent factors. Consequently, their concentration is highly scenario-specific, and may depart significantly from the general model developed at this stage. Nevertheless, even with this assumptions in mind, it is worth to carry out this exercise to obtain a first estimate of the impact of dust contamination on GW detector sensitivity.

\begin{figure}[!ht]  
\centering
\includegraphics[width=0.9\columnwidth]{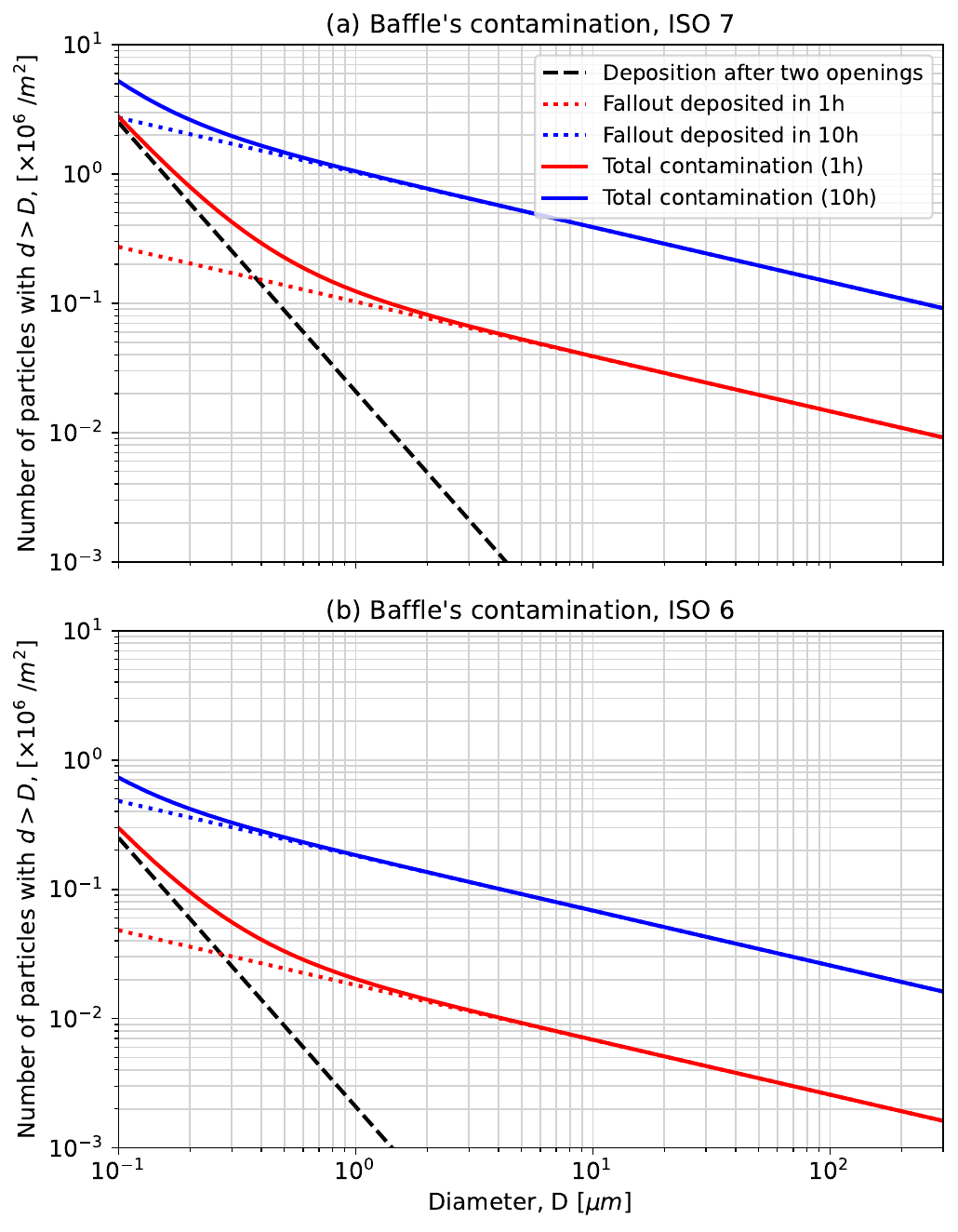}
\caption{Cumulative number of particles per unit surface expected to deposit on a ET armpipe baffle. Two scenarios are considered, with operations on baffle occurring in a ISO~7 (panel $(a)$) and ISO~6 (panel $(b)$) cleanroom. Dotted lines represent the particles that deposit during the exposure in the cleanroom, for $t=1\mathrm{h}$ (red) and $t=10\mathrm{h}$ (blue). Black dashed lines represent twice  (see Sec.\ref{sec:req} for an explanation) the contribution coming from the deposition of all particles present in the pipe section volume after this has been sealed, and is independent on $t$.}
\label{fig:contamination}
\end{figure}

The resulting cumulative number of particles deposited per unit surface area is shown as dotted lines in the top panel of Fig.~\ref{fig:contamination}, for exposure times $t =$ 1 and 10 hours. For different ISO classes or exposure times,  the numerosity scales by a factor $10^{\mathrm{ISO}}$ and linearly with $t$: the scaling is evident also by comparing the corresponding curves in the bottom panel of Fig.~\ref{fig:contamination} which shows the case expected for ISO~6 cleanroom. 

\begin{figure}[!h]
\includegraphics[scale=0.22]{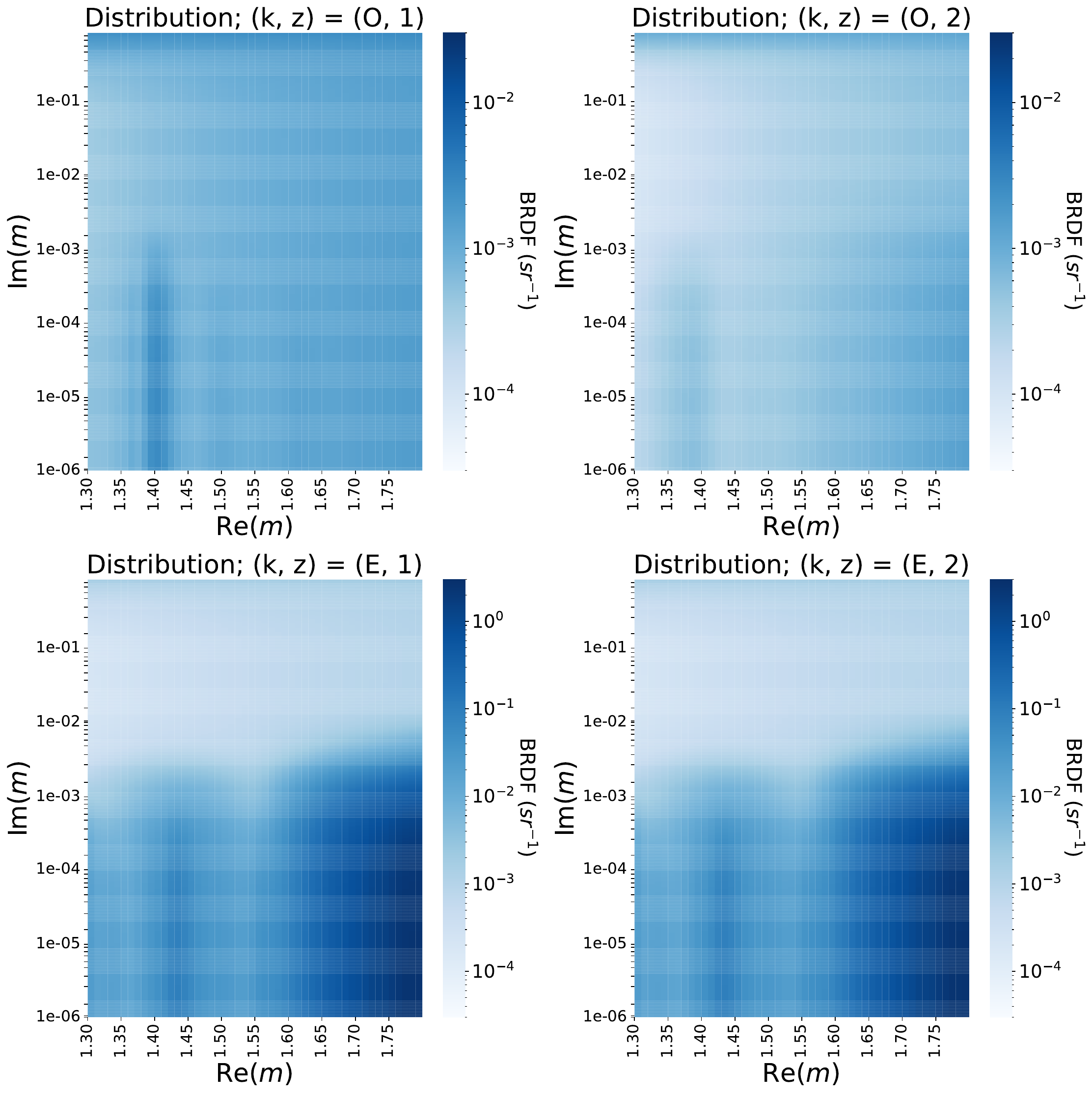}
\caption{Same as Figure~\ref{fig:brdf_vs_m_1} but for the distribution of particles (all having the same $m$) expected after horizontal exposure for $t=10\mathrm{h}$ in a ISO~7 cleanroom. For producing these maps the particle size distribution is sampled linearly with spacing $dx=0.3$ and $\lambda=1064 \mu m$. The plots in the same row share the same color scale. The upper ($k=O)$ and lower row ($k=E$) share the same dynamical range of the corresponding rows in Fig.\ref{fig:brdf_vs_m_1} and Fig.\ref{fig:brdf_vs_m_100}.}
\label{fig:brdf_vs_m_distribution}
\end{figure}

As discussed at the end of Sec.\ref{sec:ETcase}, we expect the pattern of local maxima and minima to be smoothed out when considering a distribution of particles with different sizes. This is evident from Fig.\ref{fig:brdf_vs_m_distribution} which shows the same heatmaps as in Fig.\ref{fig:brdf_vs_m_1} and Fig.\ref{fig:brdf_vs_m_100} but for the size and numerosity distribution expected from 10 hours of exposure in ISO~7 cleanroom. In general, the scattering contribution from the $k=E$ cases is similar or higher than for $k=O$, for $\mathrm{Im}(m) \lesssim 10^{-2}$.

\subsection{Requirements}
\label{sec:req}
The slope of measured particle size distribution in cleanrooms is often reported to depart from the ideal ISO standard \cite{hamberg}. To provide cleanliness requirements that are practical to apply and test in different scenarios, we divide the particle diameter range in sub-intervals and compute the maximum number of particles that satisfies Eq.\eqref{eq:condition} for each interval separately. In the following section, the derivation and results are presented.

\begin{figure}[!ptb]  
\includegraphics[scale=0.4]{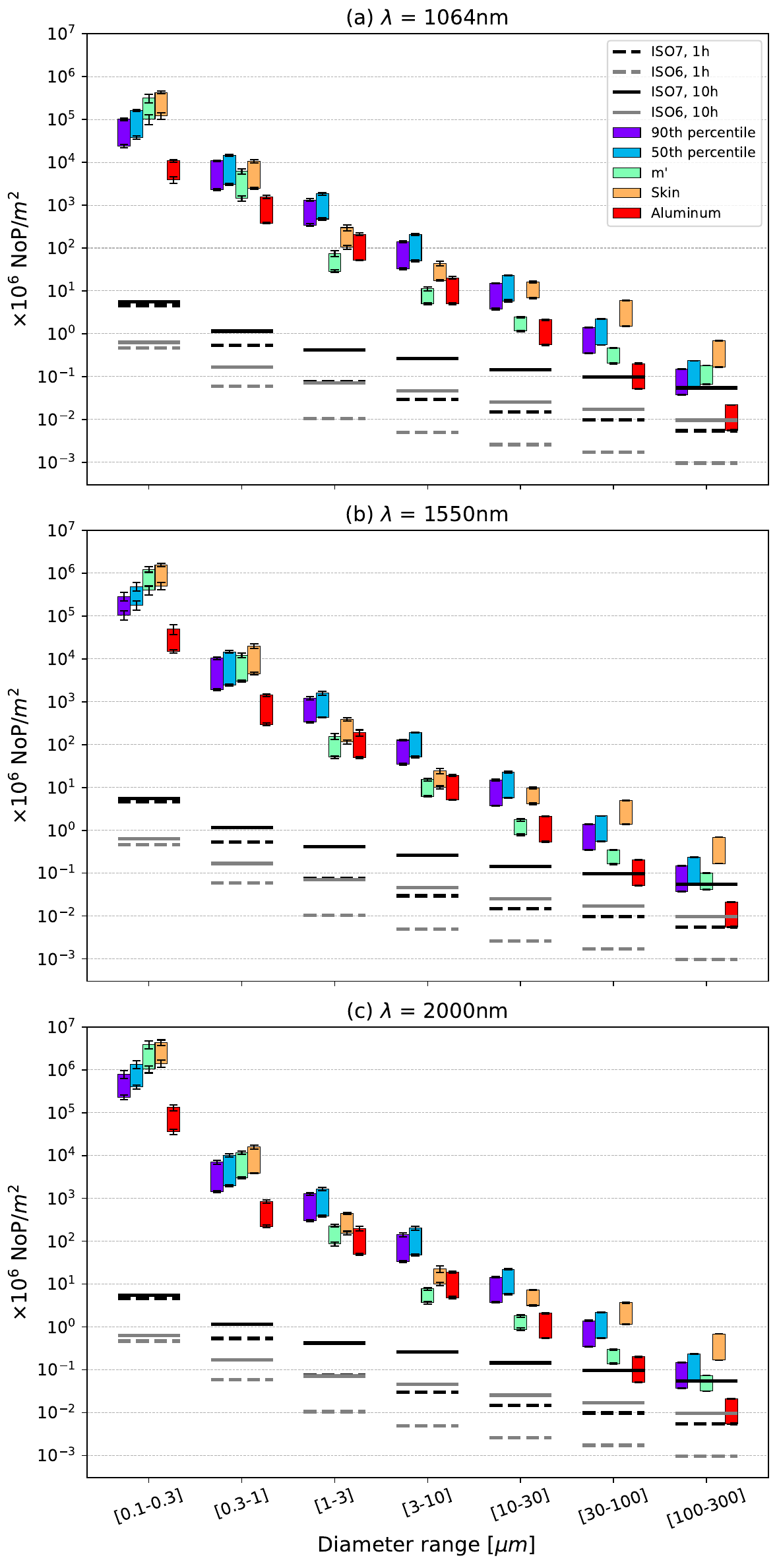}
\caption{Maximum allowed density of particles on the ET arm baffles for $\lambda = 1064$nm (a), $1550$nm (b) and $2000$nm (c), for each diameter sub-interval $D_j$. For each rectangle, the lower and upper edges represent the maximum allowed particle density $f_j$ averaged over five simulations, assuming baffle reflectivities $R=1$ and $R=10^{-2}$, respectively.  Error bars indicate the standard deviation across the five simulations. The values of $f_j$ that satisfy Eq.\eqref{eq:condition} for $BRDF_{50}$ and $BRDF_{90}$ (see the text for the definitions) are shown in purple and blue, respectively. The densities satisfying the same criterion for specific refractive indices --- $m=m^*$, and values for skin and aluminum --- are shown in cyan, orange and red, respectively. Also shown is the estimated density of particles deposited on the baffles during the installation of the pipe sectors and baffles, as described in Sec.\ref{sec:req}. Two cleanroom classes are considered: ISO~7 (black) and ISO~6 (gray). Solid lines correspond to a total exposure time of 1 hour, while dashed lines correspond to 10 hours. These curves correspond to the data from the solid lines of Fig.\ref{fig:contamination}, integrated over the corresponding diameter range.}
\label{fig:density_req}
\end{figure}

We arbitrarily divide the diameter range $[ 0.1, 300 ] \, \mu$m, shown in Figure \ref{fig:contamination}, into seven sub-intervals, denoted as $D_j$, as indicated along the horizontal axis of Figure~\ref{fig:density_req}. Within each sub-interval we assume a uniform probability distribution, i.e. the probability density $f(D) = f_j$ is constant within each bin. For each diameter bin $D_j$, we sample randomly $N_j=\text{min}\{50; \ 10\cdot (D_j/1\mu m)\}$ values.

Since the material of the dust particles in ET is unknown, we adopt a statistical approach. For each value of the refractive index $m$ in the discrete grid described in Sec.\ref{sec:ref_index}, we compute $BRDF_{dust, m}$ at $\theta_s=-\theta_i=-\theta_B$ assuming that all particles have the same value of $m$.
We weight the probability of each value of $BRDF_{dust, m}$ accounting for the spacing of the $m$ values in the $\mathrm{Im}(m)$ axis, obtaining a probability distribution $P_j(BRDF)$ for the value of the $BRDF$ generated by particles in each diameter bin $D_j$. This computation is repeated for the two considered reflectivity values, $R=1$ and $R=10^{-2}$ (see Sec.\ref{sec:R_and_pol}) and for each wavelength of interest.

Finally, the random extraction of particle diameters inside the bins is then repeated five times, $i=1...5$, obtaining five different realization of the the $BRDF$ probability distribution for each set of parameters. The resulting  distributions $P_{j,R,\lambda, i}(BRDF_{dust})$ --- computed for each of the seven diameter ranges $D_j$, two surface reflectivities $R$, three wavelengths $\lambda$ and five repeated simulations $i$ --- are reported and discussed in Appendix~\ref{sec:brdf_distr}.

For each $P_{j,R,\lambda, i}(BRDF_{dust})$ distribution, we identify the values $BRDF_{50}$ and $BRDF_{90}$ that mark the 50th and 90th percentiles of the weighted distribution. We then determine the maximum allowed particle density that lead to $BRDF_{50}$ and $BRDF_{90}$ satisfying Eq.\eqref{eq:condition}. The maximum allowed particle density is also computed using $BRDF_{dust}$ values obtained with fixed refractive indices: $m = m^*$ , as well as the values for skin and aluminum.

The five repetitions of random extraction of diameters for each bin $D_j$ are used to obtain an estimate of the mean and variance of these results.

These results are summarized in Fig.\ref{fig:density_req}. The particle densities reported as colored boxes represent the cleanliness requirements for the arm baffles in the ET vacuum  pipes, for the three considered wavelengths: $\lambda = 1064$~nm (top), $\lambda=1550\mathrm{nm}$ (center) and $\lambda=2000\mathrm{nm}$ (bottom). Larger particles appear to be the most problematic in terms of cleanliness, as their maximum tolerable numerosities are closer to the ones predicted for the specified cleanrooms classes and exposure time. Overall, the results are similar across all wavelengths.

It must be stressed that the maximum tolerable particles densities, indicated as colored boxes in Fig.\ref{fig:density_req}, do not assume any specific particle deposition model. Therefore, these results are not affected by the assumptions and limitations discussed in Sec.\ref{sec:dist}: such assumptions are only needed when translating those limits into cleanroom specifications.

\subsection{Guidelines}
\label{sec:guidelines}
We use these results to analyze cleanliness issues related to the assembly and installation of the ET beam pipes, which will be a major challenge in terms of cost and engineering. The vacuum pipes of all GW interferometric detectors, including ET, are composed in modules. According to the current design, about 8000 modules, each $\sim 15$m long \cite{Grado2023}, are connected to realize the ET arm pipe in full scale. A large number of baffles ($\sim 100-300$) need to be installed in each ET arm \cite{PhysRevD.108.102001}.

\begin{figure}[!h]
\centering
\includegraphics[width=0.8\linewidth]{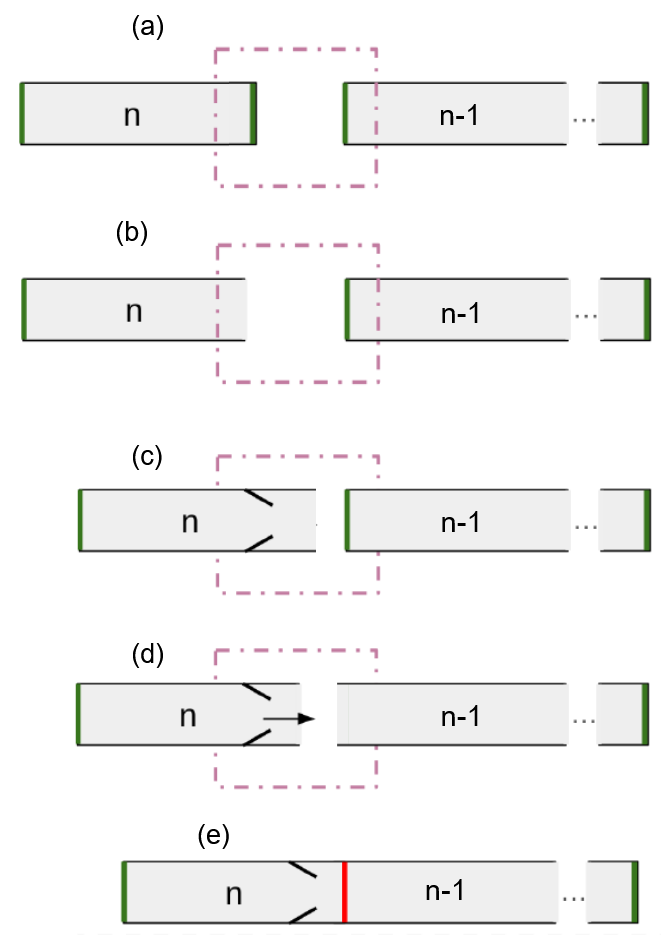}
\caption{
Generic installation sequence for the ET arm pipes with baffles. Tube modules are shown as gray rectangles. The module to be installed is labeled as $n$, while the last previously installed module is labeled as $n-1$. Seals at the section ends are indicated by green solid lines. The sequence proceeds from top to bottom: (a) a clean, sealed module is positioned adjacent to the previously installed beam pipe section (also sealed), and the end to be welded is placed in a clean environment (represented by the dotted-dashed rectangle); (b) the seal of the end to be joined is opened; (c) (if required) a baffle (black tilted lines) is installed in the new section; (d) the seal of the last installed section is opened and the new section is brought into contact with the latter; (e) the clean tent is removed, and the two sections are welded together (joint indicated by a red line).}
\label{fig:installation}
\end{figure}

Our goal is to identify situations that are most likely to  cause dust to enter and accumulate, to estimate the corresponding surface contamination, and to give guidelines that can help optimize the construction procedure. While the details of the construction and transportation strategies for the single sections are not yet defined, we consider here a generic and yet representative scenario for the final stage of assembly, as represented in  Figure~\ref{fig:installation}.
We start by considering a to-be-installed clean pipe module (i.e. produced, cleaned and handled ensuring the requirements of Figure \ref{fig:density_req} are met), sealed at both ends, which has been brought in the proximity of an already partially assembled (and temporarily sealed) part of the ET beam pipe. We consider the air in both the new section and the partial ET arm pipe to be almost pristine, as most of the suspended particulate will have had time to deposit, leaving the air ``clean''. We also assume that a clean environment is established such that it encloses the openings of the pipes to be joined together (a movable clean tent can be imagined for the sake of concreteness).
The new section is then opened at one end to install a baffle (considered also to be pristine, or at least to meet requirements from Figure \ref{fig:density_req}); during this phase, the section is filled with air from the clean tent and the baffle is exposed to the clean environment for the time needed for installation.
Once the baffle is installed, the already assembled part of the ET beampipe is opened, and the two are are brought in contact; during this stage, we assume that the first section of the already assembled ET beampipe will also be filled with air at the cleanliness level of the tent. This is a crude but conservative approximation, both because it is not easy to imagine air currents that can propagate deep into the narrow geometry of the pipe (sealed at the other end), and because it would be relatively easy to generate a mild overpressure of clean air in the tube to prevent air from entering.
Finally, the two sections are joined (the details of the process are still to be defined \cite{ETpipe}) and the whole procedure can start over with a new section.

Based on this simple scenario and some assumptions and approximations, we can provide a coarse estimate of the cleanliness requirements for the clean environment and for the exposure time of the baffle to the clean environment.
When a seal is opened, the entire volume of the pipe is filled with new air at the cleanliness level of the tent environment. The total contamination expected for each baffle is the sum of two contributions. The first one comes from particles that start to deposit throughout the section, including  the baffle: the number of deposited particles scales with the exposure time $t$, as described in Sec.\ref{sec:dist}.
The second contribution arises after the sections are joined and sealed again. At this point, all particles still suspended in the enclosed volume will eventually deposit (with a timescale of the order of a day or less, as can be deduced by comparing, in Fig.~\ref{fig:density_req}, the total number of particles deposited by this mechanism with that deposited by 10 hours of exposure).
We assume these particles settle uniformly on the internal surface of the pipe section, meaning only a fraction will deposit on the baffles. This contribution is independent of the exposure time $t$, and is experienced twice by each section/baffle: once when it is newly installed and again when it is opened again to install the next section.
Conservatively, in this analysis we neglect geometric corrections associated with the baffles' shape, which slightly reduces the effective area exposed to falling particles, as well as the fact that half of the baffle is actually facing downwards.

Solid lines in Fig.~\ref{fig:contamination} show the cumulative surface particle density expected to deposit on a baffle from the sum of the two contributions, for an ISO~6 and ISO~7 clean tent and for exposure times $t=1$~h and $t=10$~h. In Fig.\ref{fig:density_req} we show the same data but integrated over each diameter sub-interval $D_j$: hence, for the first time, expected contamination can be compared directly with the cleanliness requirements set by SL noise considerations. Constraints are tighter for larger particles: the density of particles that would deposit during the installation process for specific combinations of ISO and $t$ is comparable to (if not larger than) the maximum allowed particle density, for some $m$ values. The contamination by the mid-size to large particles, causing most of the SL, is dominated by the exposure occurring while the seal is open: this is proportional to the exposure time $t$ (see also Fig.\ref{fig:contamination}), which is thus a critical parameter.

%\textcolor{gray}{Despite the simplifying assumptions, this model suggests that SL noise will be likely dominated by large particles, and that operation in ISO 7 for several hours ($\gtrsim $ 10 hours) may not provide a large enough safety margin to prevent the SL noise from the contaminated baffles becoming detrimental to the sensitivity of ET.}

Our model suggests that SL noise will be likely dominated by large particles. Since our model is based on an arbitrary extension of the ISO standard to larger particles, a better characterization of the particle surface concentration will be needed to obtain more solid cleanliness requirements.

%%\textcolor{gray}{It has to be noted that the ISO particle distribution is adopted as a first-order approximation to derive illustrative cleanliness requirements. In particular, ISO cleanliness requirements are specified for particles up to $5\mathrm{\mu m}$ in size, and for larger particles the distribution is obtained by extrapolation. Although this extrapolation has been experimentally validated up to $25\mathrm{\mu m}$ (see Ref.\cite{moscatello_phd}, Sec.5.1), its applicability to larger particles is more uncertain. However, deposition of larger particles depends strongly on the specific cleanroom layout, nearby activities, handling procedures, personnel movements, and other process-dependent factors. Consequently, they are highly scenario-specific and cannot be reliably predicted within a general model developed at this stage. For this reason, cleanliness requirements derived for the largest particle sizes should be interpreted as indicative estimates, to be refined once the assembly procedures and contamination sources are better characterized.}

\section{Conclusions}
\label{sec:concl}
In this paper we describe how a distribution of particles deposited on a reflective surface scatters light and explain the different mechanisms involved. We show how the scattered light power distributes over the solid angle and discuss how this changes with respect to light polarization and particle index of refraction and size. 

We study the specific case of the baffles of the ET vacuum arm pipes and we find that the most harmful particles have diameters $D>\lambda$ and refractive index $m$ with high Re($m$) and low Im($m$).

We set the cleanliness requirement for the ET arm baffles at the three wavelengths under consideration for the observatory, showing that for both ET-HF and ET-LF the cleanliness requirements are similar.

We give guidelines from the cleanliness viewpoint for the assembly of the ET vacuum pipes with baffles and show that working in an ISO7 clean room may not be enough if the installation procedures are prolonged in time. After the installation procedures are established in more details, this study should be repeated to provide tailored constraints.

To define cleanliness requirements on the inner surface of the pipe and complete the cleanliness analysis for the ET arms, the next step in this study is to investigate the stray light noise caused by dust particles falling under gravity in vacuum and crossing the laser beam of the ET interferometer arms. A further step would be to assess the cleanliness requirements for the vacuum towers that host the core and suspended optics of the interferometers: the work presented here provides a foundation for that analysis.

In summary, this study has demonstrated how insufficient cleanliness can contribute to the generation of stray light noise in the Einstein Telescope. More broadly, high-sensitivity experiments and devices that rely on optical technologies are also affected by stray light, and maintaining adequate cleanliness is often a key strategy to mitigate this issue. Our work provides a foundation for scientifically quantifying what adequate cleanliness truly means in these cases.

\appendix

\section{Coherent summation of terms in $BRDF_{dust}$}
\label{sec:brdf_coherent}
The expressions for the $BRDF$ in Eq.s \eqref{eq:brdf_dust_SS_incoherent} and \eqref{eq:brdf_dust_PP_incoherent} are evaluated as the incoherent sum of the odd- and even-parity scattering contributions. In this appendix, we derive the equations for the $BRDF$ in the coherent case and discuss why, in realistic scenarios, the incoherent approach is justified.

\begin{figure}[htpb]
    \centering
    \includegraphics[width=0.95\columnwidth]{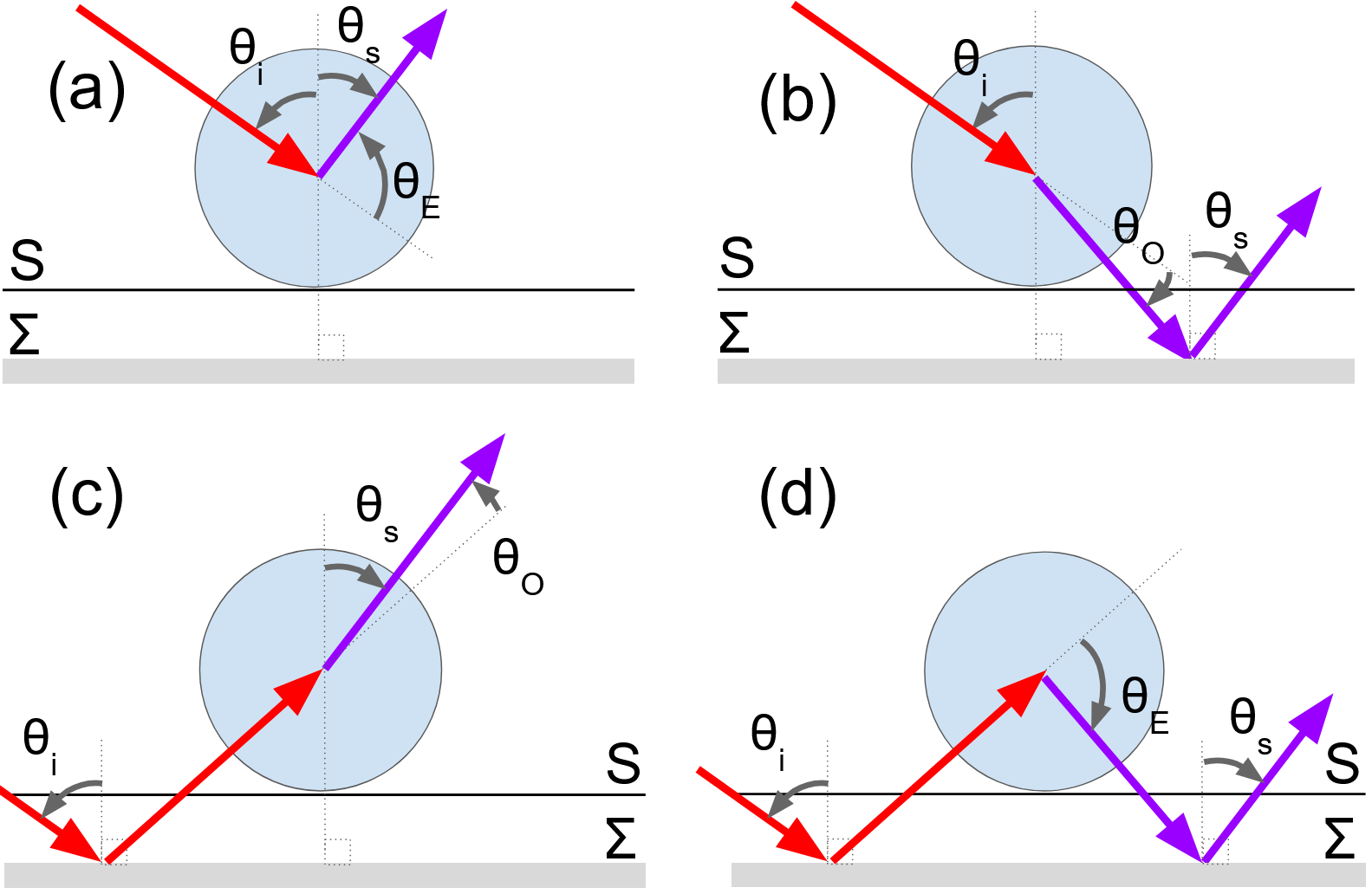}
    \caption{Same as Fig.\ref{fig:3processes_h}, but the optical (light-gray tick line) and the contact (dark-gray thin line) surfaces are separated.}
    \label{fig:4mech_coherent}
\end{figure}

Let us consider a particle deposited -- i.e. in physical contact -- on a real surface $S$ which is located at a distance $h$ from the effective reflection surface $\Sigma$; indicated in Fig.\ref{fig:4mech_coherent}. The two surfaces are in general distinct, especially in optical components, where coatings (anti-reflective, highly-reflective or even just protective) are often used. $h$ depends on the nature and composition of the surface, and on the wavelength of light considered.

In this case the scattering mechanisms in Fig.\ref{fig:3processes_h} are modified as in Fig.~\ref{fig:4mech_coherent}.

\begin{figure}[!h]
    \centering
    \includegraphics[width=0.6\linewidth]{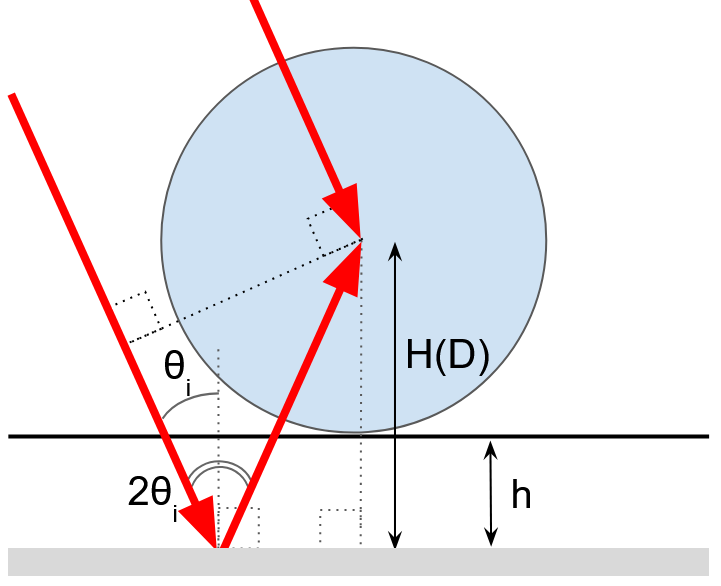} 
    \caption{Schematic representation of the geometry for two light rays: one arriving directly on the particle and the other being reflected by the surface toward the particle. As for fig.\ref{fig:3processes_h}, the horizontal black line represents the physical surface, while the thicker gray line represent the effective optical surface.}
    \label{fig:path_length}
\end{figure}

When applying Mie theory to an isolated particle, light hits the particle directly and leaves in the direction of interest, with the only reference point for the phase of both fields being the center of the particle. When considering alternative paths for the light, either incoming or outgoing, it is thus necessary to calculate the additional phase accumulated by the light along these paths, which in the case under consideration depends on the distance $H$ between the particle center and $\Sigma$.
In turn, $H$ depends on the particle diameter $D$, since it is the sum of the separation between the effective and real surfaces, $h$, and the particle radius. Thus, 
\begin{equation}
H(D) = h + \tfrac{D}{2}.
\end{equation}

Fig.~\ref{fig:path_length} illustrates the geometry to calculate the path-length differences for light arriving at the particle after reflection on the surface $\Sigma$ (the case of reflection after scattering by the particle is analogous). The total optical path length difference is obtained as the sum of two contributions: the additional path length before reaching the surface, $\Delta l'$, and the path length between the reflection and the particle, $\Delta l''$:
\begin{equation}
    \begin{split}
        \Delta l'' &= \frac{H(D)}{\cos(\theta_i)} = \frac{h + D/2}{\cos(\theta_i)}, \\
        \Delta l'  &= \Delta l'' \cdot \cos(2\theta_i) =  \left(h+\frac{D}{2}\right) \frac{\cos(2\theta_i)}{\cos(\theta_i)}.
    \end{split}
\end{equation}
Therefore, the phase difference of the two light rays arriving to the particles is:
\begin{equation}
    \Delta \phi (\theta_i, h, D) = \frac{2\pi}{\lambda} \left(h+\frac{D}{2}\right) \left[  \frac{\cos(2\theta_i)+1}{\cos(\theta_i)} \right].
    \label{eq:delta_phi}
\end{equation}
In the case in which light is first scattered by the particle and then reflected by the surface, the phase difference can also be calculated using Eq.\ref{eq:delta_phi}, provided that $\theta_i$ is replaced with $\theta_s$. Finally, the coherent expression of $BRDF_{dust}$ --- which accounts for the summation of all scattering contributions --- can be written as:
\begin{equation}
\begin{split}
BRDF_{dust}^{SS}(\lambda, m, D, &\theta_i, \theta_s) = \frac{\lambda^2}{4\pi^2 \cos(\theta_s)} \int f(x) ...\\
\times | (F_{a}^{S}& + F_{d}^{S}) S_{1}(x, m, \theta_{E})\\
+ (F_{b}^{S}& + F_{c}^{S}) S_{1}(x, m, \theta_{O})|^2
\end{split}
\label{eq:brdf_dust_SS_coherent}
\end{equation}
\begin{equation}
\begin{split}
BRDF_{dust}^{PP}(\lambda, m, D, &\theta_i, \theta_s) = \frac{\lambda^2}{4\pi^2 \cos(\theta_s)} \int f(x) ... \\ 
\times | (F_{a}^{P}& + F_{d}^{P}) S_{2}(x, m, \theta_{E})\\
+ (F_{b}^{P}& + F_{c}^{P}) S_{2}(x, m, \theta_{O})|^2
\end{split}
\label{eq:brdf_dust_PP_coherent}
\end{equation}
The complex coefficients $F^{S,P}_{a,b,c,d}$ account for the reflectivities and phase shifts in the respective cases illustrated in Fig.~\ref{fig:4mech_coherent}. As light is passing from air/vacuum ($n_1=1$) to a medium with $n_2 > n_1$, at each interaction with the surface a $\pi$ phase is added upon reflection from the optical plane, while transmission in the medium between the contact and optical surface does not add a phase shift, nor the passage from the medium to air/vacuum. Furthermore, the distance $h$ is, in general, dependent on the polarization; we express with $h_S$ and $h_P$ the distance for the S- and P-polarized light respectively. Finally, the coefficients are written as:
\begin{equation}
    \begin{split}
        F_{a}^{S,P} & = 1 \\
        F_{b}^{S,P} & = \sqrt{R_{S,P}(\theta_{s})} \exp\left[ i \Delta \phi (\theta_{s}, h_{S,P}, D) \right]\\
        F_{c}^{S,P} & = \sqrt{R_{S,P}(\theta_{i})} \exp\left[ i \Delta \phi (\theta_{i}, h_{S,P}, D) \right] \\
        F_{d}^{S,P} & = \sqrt{R_{S,P}(\theta_{s})R_{S,P}(\theta_{i})} \times \\
        &\quad \exp\left[ i \Delta \phi (\theta_{s}, h_{S,P}, D) + i \Delta \phi (\theta_{i}, h_{S,P}, D) \right]\\
    \end{split}
    \label{eq:brfd_factors}
\end{equation}

While the model is valid in general, for the specific case of dust contamination on ET baffles the incoherent summation (Eq.s \eqref{eq:brdf_dust_SS_incoherent} and \eqref{eq:brdf_dust_PP_incoherent}) is preferred because, as discussed in the main text: (i) particles are not spherical and (ii) optical properties of the baffles, therefore the choice of $h$ would be arbitrary.

%Firstly, depending on the specific $h$ value, different degrees of constructive or destructive interference occur at the level of the single particle, so that the final $BRDF$ estimate depends heavily on the value of $h$. At this stage the optical properties of the baffles are not defined yet (further details in Sec.\ref{sec:et_req}): in particular, the value of $h$ is not known, and any assumption would be arbitrary.

%Secondly, even if $h$ were known in advance, real particles are not spherical: even when they have the same circular equivalent area, the distance between their center and the real surface will depend on shape and orientation. Averaging over different shapes and orientation is equivalent to perform  the incoherent summation of all their scattering contributions.

\section{Polarization of SL from dust particles on ET baffles}
\label{sec:pol}
The polarization of (the vast majority of) light circulating in the ET arm pipes is linear, and this is true also for light reaching the baffles (including light scattered by a TM), as low-angle scattering from the TMs is expected to preserve polarization.
The rest of the geometry has cylindrical symmetry; for each event of particle scattering, the scattering plane --- defined by the point on the TM where the photon originates, the dust particle on the baffle surface, and the point of the TM reached by the scattered photon --- is oriented arbitrarily. Thus, from the point of view of the individual particle scattering events, incident light is equally distributed between polarization 1 and 2 used in Eq.\ref{eq:brdf_dust}.
However, when light scattered by deposited particles is projected in the same basis of the light circulating in the arm, averaging over the rotation angles of the scattering plane returns a non-null component only along the polarization of the incident light; the perpendicular polarization averages out.

To prove this, we define the $(\hat{x}, \hat{y})$ basis to describe the polarization of the field circulating in the arm: we assume it is aligned along $\hat{x}$ (the derivation is identical for the orthogonal case). For each scattering event, let $\eta$ be the angle between $\hat{x}$ and the scattering plane. From Eq.\ref{eq:mie_general}, the scattered field in the $(\hat{x}, \hat{y})$ basis is:

\begin{equation}
\begin{pmatrix}
E^{s}_x(\eta) \\
E^{s}_y(\eta)
\end{pmatrix}
= \text{Rot}(-\eta)
\begin{pmatrix}
S_1 & 0 \\
0 & S_2
\end{pmatrix}
\text{Rot}(\eta)
\begin{pmatrix}
E^{i}_x \\
0
\end{pmatrix}.
\label{eq:rotation_pol}
\end{equation}
where $\mathrm{Rot}(\eta)$ is the rotation matrix between the detector frame and the scattering plane:
\begin{equation}
\text{Rot}(\eta) = 
\begin{pmatrix}
\cos\eta &  -\sin\eta \\
\sin\eta & \cos\eta
\end{pmatrix}.
\end{equation}
The propagating factor is independent of the polarization basis rotation so it is taken as unitary. By explicitly performing the calculations, Eq.\ref{eq:rotation_pol} becomes:
\begin{equation}
\begin{split}
\begin{pmatrix}
E^{s}_x(\eta) \\
E^{s}_y(\eta)
\end{pmatrix}
= E^{i}_x
\begin{pmatrix}
\big(S_1 \cos^2\eta + S_2 \sin^2\eta \big) \\
-\big(S_1 - S_2\big)\sin\eta \cos\eta 
\end{pmatrix}
\end{split}
\label{eq:none}
\end{equation}
If the particles are randomly distributed on each baffle and because of the cylindrical symmetry of the problem, $\eta$ is uniformly distributed over $[0, 2\pi]$. Therefore, by averaging over $\eta$ the scattered field averages to
\begin{equation}
\begin{pmatrix}
\langle E^{s}_x \rangle \\
\langle E^{s}_y \rangle
\end{pmatrix}
= 
\begin{pmatrix}
\frac{1}{2} \left( S_1 + S_2\right) E^{i}_x  \\
0
\end{pmatrix},
\end{equation}
where we see that it is aligned with the input polarization. This allows us to compute the Mie scattering in the unpolarized light case (in the scattering plane basis).

\section{Weighted distributions of the $BRDF$}
\label{sec:brdf_distr}
In Sec.~\ref{sec:et_req}, the derivation of cleanliness limits for the ET baffles (Fig.~\ref{fig:density_req}) is based on a series of Monte Carlo simulations. The results shown there are obtained as follows. For each $j$-th diameter range:
	\begin{enumerate}
		\item for five independent runs ($i=1,...,5$):  
			\begin{enumerate}
				\item $N_{j} =\text{min}\{50; \ 10\cdot (D_j/1\mu m)\}$ diameters values are randomly sampled with uniform probability inside the bin.
				\item All the $N_{j}$ particles are assigned the same $m$, for each $m$ value in the grid (see Sec. \ref{sec:et_req}). The resulting $BRDF(m)_j^i$ is computed.
                \item Each $BRDF(m)_j^i$ is assigned a weight based in the spacing of the discrete $\mathrm{Im}(m)$ values in the grid (spacing of $\mathrm{Re}(m)$ values is uniform, so it does not matter). This weight only accounts for non-uniform sampling, while the probability of $m$ is considered uniform over the whole domain considered.
                \item From the weighted $BRDF(m)_j^i$, a probability distribution  $P(BRDF_j^i)$ is obtained.
				\item From the $BRDF_j^i$ distribution, we determine the 50th and 90th percentiles. In summary we obtain the 50th, 90th percentiles of the $BRDF$ distribution over the whole ensemble of $m$ values, for $N_j$ randomly sampled particles in the diameter range $j$: ${BRDF_{50}}_{j}^{i}$, ${BRDF_{90}}_{j}^{i}$.
			\end{enumerate}
		\item The average over $i$ of the percentiles ${BRDF_{50}}_{j}^i$ and ${BRDF_{90}}_{j}^i$ and their standard deviations are computed. This allows to estimate and average fluctuations caused by specific diameter samplings.
		\item From the mean percentile with its standard deviation determined at previous step, we compute the average and standard deviation of the particles numerosities that satisfies Eq.\ref{eq:condition}. The results are reported as the upper and lower edges of the boxes, and corresponding error bars, of the candle plot in Fig.\ref{fig:density_req}
	\end{enumerate}
The process is then repeated for the other $m$ values under study: $m^{*}, m_{Al}$ and $m_{skin}$. In those cases, however, on step (2.b) the $BRDF$ is computed only once, for the specific $m$ value. Therefore, the 50th and 90th percentiles are not computed: the limit numerosities are directly computed instead and the average and standard deviations are computed over the five different realizations only.

\begin{figure}[!ht]  
\centering
\includegraphics[width=0.9\columnwidth]{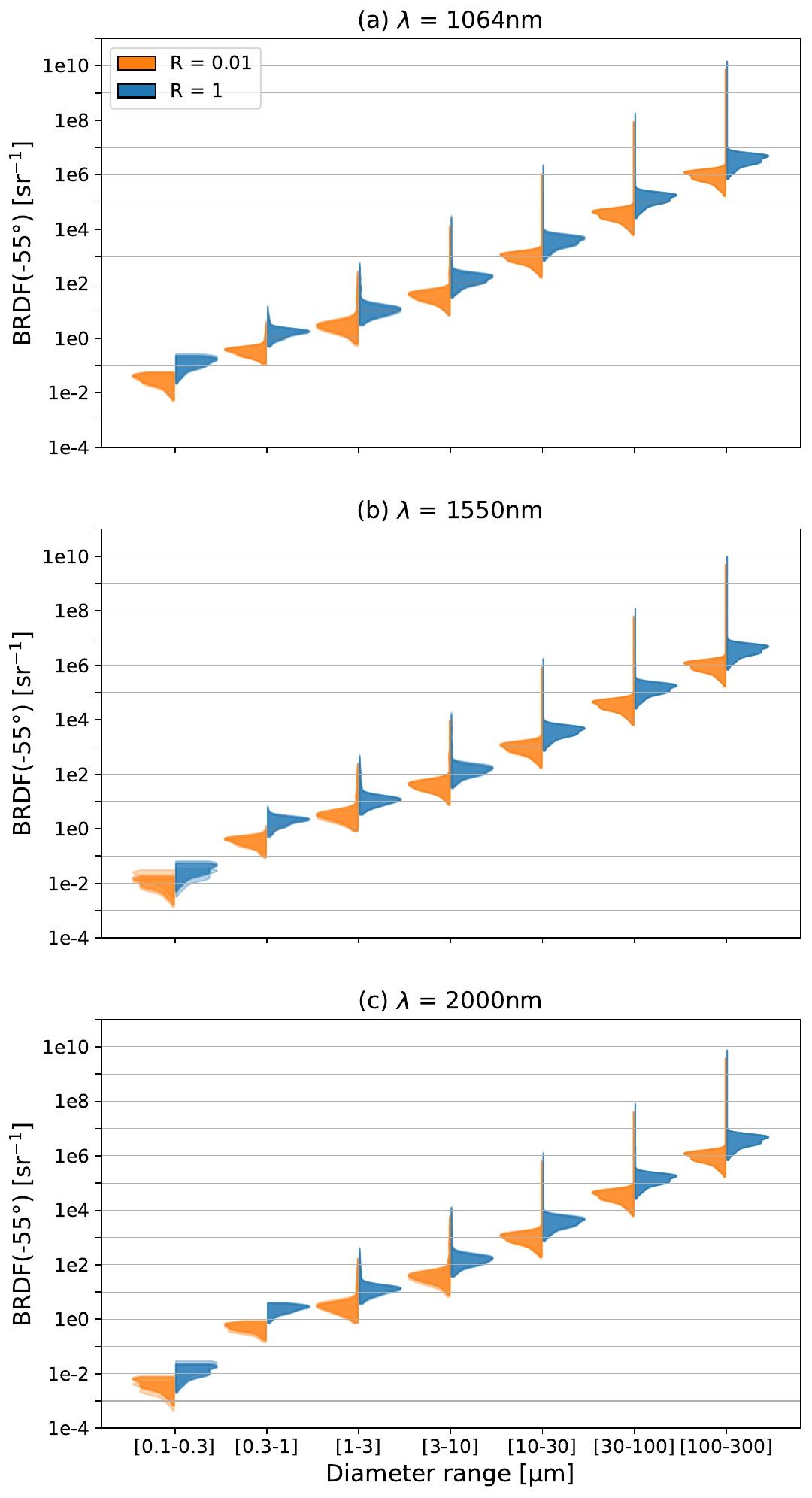}
\caption{Violin plots of the $BRDF$ values (at $\theta_s=-\theta_i=-\theta_B$) simulated with the Monte Carlo analysis described in Sec.~\ref{sec:et_req}, for three wavelengths: $\lambda = 1064~\mathrm{nm}$, $1550~\mathrm{nm}$, and $2000~\mathrm{nm}$, as indicated in the titles. Each violin shows the distribution of $BRDF$ values for high ($R=1$, blue, right) and low ($R=0.01$, orange, left) surface reflectivity. Shaded areas correspond to the distributions obtained from five independent simulations per case. $BRDF$ values are weighted by the bin widths of $\mathrm{Im}(m)$.}
\label{fig:violin}
\end{figure}

Figure~\ref{fig:violin} presents violin plots showing the weighted $BRDF$ distributions for different particle diameter ranges, surface reflectivities, and light wavelengths. The shaded regions represent the results from all five simulations for each configuration. While minor variations are observed in the smaller diameter bins, the distributions converge as the particle size increases, indicating reduced sensitivity to the random sampling in this regime.

Notably, the upper tail of the $BRDF$ distribution, particularly for larger particles, spans several orders of magnitude. This upper tail is caused by low $\mathrm{Im}(m)$ values. These high-$BRDF$ outliers, although physically possible, represent statistically rare scenarios. For instance, in the largest diameter range, the $BRDF$ spans more than four orders of magnitude across the sampled $m$ space. As a result, a purely conservative approach based on the maximum $BRDF$ would lead to overly stringent cleanliness requirements, significantly exceeding practical contamination control standards.
To avoid this, a percentile-based approach is adopted (as discussed here and in Sec.\ref{sec:et_req}), allowing the exclusion of extreme but rare optical responses and yielding a more realistic and actionable threshold for particulate contamination control.

% Create the reference section using BibTeX:
\bibliography{ET_deposited_dust_bib}

@article{Ottaway2012,
author = {David J Ottaway and Peter Fritschel and Samuel J. Waldman},
journal = {Opt. Express},
number = {8},
pages = {8329--8336},
publisher = {Optica Publishing Group},
title = {Impact of upconverted scattered light on advanced interferometric gravitational wave detectors},
volume = {20},
month = {Apr},
year = {2012},
url = {https://opg.optica.org/oe/abstract.cfm?URI=oe-20-8-8329},
doi = {10.1364/OE.20.008329},
}

@article{Shettle79,
	author = {Shettle, Eric and Fenn, Robert},
	year = {1979},
	month = {09},
	pages = {94},
	title = {Models for the Aerosols of the Lower Atmosphere and the Effects of Humidity 	Variations on their Optical Properties},
	journal = {Environ. Res.}
}

@article{LIGO:2020zwl,
    author = "Soni, S. and others",
    collaboration = "LIGO",
    title = "{Reducing scattered light in LIGO's third observing run}",
    eprint = "2007.14876",
    archivePrefix = "arXiv",
    primaryClass = "astro-ph.IM",
    reportNumber = "LIGO-P2000172",
    doi = "10.1088/1361-6382/abc906",
    journal = "Class. Quant. Grav.",
    volume = "38",
    number = "2",
    pages = "025016",
    year = "2020"
}

@techreport{Thorne1989,
     title = {{Light Scattering and Proposed Baffle Configuration for the LIGO}},
     author = {Kip Thorne},
     year = {1989},
	 NUMBER = {No. LIGO-T890017-x0}
}

@article{Punturo_2010,
doi = {10.1088/0264-9381/27/19/194002},
url = {https://dx.doi.org/10.1088/0264-9381/27/19/194002},
year = {2010},
month = {sep},
publisher = {},
volume = {27},
number = {19},
pages = {194002},
author = "{M. Punturo {\it et al.}}",
title = {The Einstein Telescope: a third-generation gravitational wave observatory},
journal = {Classical and Quantum Gravity},
}

@misc{evans2021,
      title={A Horizon Study for Cosmic Explorer: Science, Observatories, and Community}, 
      author="{M. Evans {\it et al.}}",
      year={2021},
      eprint={2109.09882},
      archivePrefix={arXiv},
      primaryClass={astro-ph.IM},
      url={https://arxiv.org/abs/2109.09882}, 
}

@techreport{FlanaganThorne,
	title ={{Light scattering and baffle configuration for LIGO}},
	author={E. Flanagan and K. Thorne},
	number= {No. LIGO-T950101-00}, 
	year= 1995
}

@article{PhysRevD.108.102001,
  title = {Study of scattered light in the main arms of the Einstein Telescope gravitational wave detector},
  author = "{Andr\'es-Carcasona {\it et al.}}",
  journal = {Phys. Rev. D},
  volume = {108},
  issue = {10},
  pages = {102001},
  numpages = {16},
  year = {2023},
  month = {Nov},
  publisher = {American Physical Society},
  doi = {10.1103/PhysRevD.108.102001},
  url = {https://link.aps.org/doi/10.1103/PhysRevD.108.102001}
}

@article{PhysRevX.13.041039,
  title = {{GWTC-3: Compact Binary Coalescences Observed by LIGO and Virgo during the Second Part of the Third Observing Run}},
  author = {{Abbott, R. {\it et al.}}},
  collaboration = {LIGO Scientific Collaboration, Virgo Collaboration, and KAGRA Collaboration},
  journal = {Phys. Rev. X},
  volume = {13},
  issue = {4},
  pages = {041039},
  numpages = {89},
  year = {2023},
  month = {Dec},
  publisher = {American Physical Society},
  doi = {10.1103/PhysRevX.13.041039},
  url = {https://link.aps.org/doi/10.1103/PhysRevX.13.041039}
}

@article{Huang2021,
author = {Huang, Yue and Adebiyi, Adeyemi A. and Formenti, Paola and Kok, Jasper F.},
title = {Linking the Different Diameter Types of Aspherical Desert Dust Indicates That Models Underestimate Coarse Dust Emission},
journal = {Geophysical Research Letters},
volume = {48},
number = {6},
pages = {e2020GL092054},
doi = {https://doi.org/10.1029/2020GL092054},
url = {https://agupubs.onlinelibrary.wiley.com/doi/abs/10.1029/2020GL092054},
year = {2021}
}

@Article{DiBiagio_dust,
AUTHOR = {Di Biagio, C. {\it et al.}},
TITLE = {Complex refractive indices and single-scattering albedo of global dust
aerosols in the shortwave spectrum and relationship to size and iron content},
JOURNAL = {Atmospheric Chemistry and Physics},
VOLUME = {19},
YEAR = {2019},
NUMBER = {24},
PAGES = {15503--15531},
URL = {https://acp.copernicus.org/articles/19/15503/2019/},
DOI = {10.5194/acp-19-15503-2019}
}

@techreport{ETpipe,
	title ={{ET-PP Deliverable 6.2 [Vacuum Pipe Design]}},
	author={{P. Chiggiato  {\it et al.}}},
	number= {ET-0005A-25}, 
	year= {2025}
}

@article{WhyteAgricola,
author = "{W. Whyte {\it et al.}}",
year = {2016},
month = {04},
pages = {4-10},
title = {Airborne particle deposition in cleanrooms: Calculation of product contamination and required cleanroom class},
journal = {Clean Air and Containment Review}
}

@article{spyak4,
author = "{Paul R. Spyak and William L. Wolfe}",
title = {{Scatter from particulate-contaminated mirrors. part 4: properties of scatter from dust for visible to far-infrared wavelengths}},
volume = {31},
journal = {Optical Engineering},
number = {8},
publisher = {SPIE},
pages = {1775 -- 1784},
year = {1992},
doi = {10.1117/12.58711},
URL = {https://doi.org/10.1117/12.58711}
}

@software{miepython,
  author  = {Prahl, Scott},
  title   = {{miepython}: A Python library for Mie scattering calculations},
  url     = {https://github.com/scottprahl/miepython},
  doi     = {10.5281/zenodo.7949263},
  year    = {2025},
  version = {latest}
}

@book{Fest,
  title     = "Stray Light Analysis and Control",
  author    = "Fest, Eric C.",
  year      = 2013,
  publisher = "SPIE Press"
}

@article{Grado2023,
    author = {A. Grado  {\it et al.}},
    title = "{Ultra high vacuum beam pipe of the Einstein Telescope project: Challenges and perspectives}",
    journal = {Journal of Vacuum Science $\&$ Technology B},
    volume = {41},
    number = {2},
    pages = {024201},
    year = {2023},
    month = {02},
    issn = {2166-2746},
    doi = {10.1116/6.0002323},
    url = {https://doi.org/10.1116/6.0002323},
    eprint = {https://pubs.aip.org/avs/jvb/article-pdf/doi/10.1116/6.0002323/16778350/024201\_1\_online.pdf},
}

@article{skin,
	author={Ding H. {\it et al.}},
	title="{Refractive indices of human skin tissues at eight wavelengths and estimated dispersion relations between 300 and 1600 nm}",
	journal= {Phys Med Biol.},
	year= {2006},
	volume={51},
	number={6},
	pages={1479-89},
	doi= {10.1088/0031-9155/51/6/008}
}

@article{material_db,
	author={M. N. Polyanskiy},
	title={{Refractiveindex.info database of optical constants}}, 
	journal={Sci. Data},
	volume = 11, 
	number =94,
	year = {2024},
	doi={10.1038/s41597-023-02898-2},
}

@article{Ordal:88,
	author = "{Mark A. Ordal {\it et al.}}",
	journal = {Appl. Opt.},
	number = {6},
	pages = {1203--1209},
	publisher = {Optica Publishing Group},
	title = {Optical properties of Al, Fe, Ti, Ta, W, and Mo at submillimeter wavelengths},
	volume = {27},
	month = {Mar},
	year = {1988},
	doi = {10.1364/AO.27.001203},
}

@article{spyak1,
author = "{Paul R. Spyak and William L. Wolfe}",
title = {{Scatter from particulate-contaminated mirrors. part 1: theory and experiment for polystyrene spheres and $\lambda$=0.6328 $\mu$m}},
volume = {31},
journal = {Optical Engineering},
number = {8},
publisher = {SPIE},
pages = {1746 -- 1756},
year = {1992},
doi = {10.1117/12.58708},
URL = {https://doi.org/10.1117/12.58708}
}

@article{Degallaix:19,
author = "{J. Degallaix  {\it et al.}}",
journal = {J. Opt. Soc. Am. A},
number = {11},
pages = {C85--C94},
publisher = {Optica Publishing Group},
title = {Large and extremely low loss: the unique challenges of gravitational wave mirrors},
volume = {36},
month = {Nov},
year = {2019},
url = {https://opg.optica.org/josaa/abstract.cfm?URI=josaa-36-11-C85},
doi = {10.1364/JOSAA.36.000C85},
}

@techreport{ISO14644,
type = {Standard},
key = {ISO 14644-1:2015},
year = {2015},
title = {{Cleanrooms and associated controlled environments — Part 1: Classification of air cleanliness by particle concentration}},
volume = {2015},
address = {Geneva, CH},
institution = {International Organization for Standardization}
}

@book{BohrenHuffman,
author={Craig F. Bohren, Donald R. Huffman},
publisher = {John Wiley $\&$ Sons, Ltd},
isbn = {9783527618156},
title = {Absorption and Scattering of Light by Small Particles},
doi = {https://doi.org/10.1002/9783527618156},
url = {https://onlinelibrary.wiley.com/doi/abs/10.1002/9783527618156},
year = {1998},
}

@techreport{LIGO_report_R,
author = "{A.Ananyeva, C.Torrie}",
type = "{Technical Report LIGO-
T1700128}",
year = {2018},
title = {{Infrared Reflectance and Scatter Tests for Potential Scattered Light Mitigation Coatings ($R^2$ vs. BRDF)}},
institution = {LIGO}
}

@techreport{hamberg,
  author       = {Hamberg, O and Shon, E M},
  title        = {Particle size distribution on surfaces in clean rooms. Final technical report September 1983-February 1984},
  institution  = {Aerospace Corp., El Segundo, CA (USA). Vehicle Engineering Div.},
  url          = {https://www.osti.gov/biblio/6150472},
  place        = {United States},
  year         = {1984},
  month        = {04}}

@inproceedings{10.1117/12.2047327,
author = {Margot H Phelps and Kaitlin E. Gushwa and Calum I. Torrie},
title = {{Optical contamination control in the Advanced LIGO ultra-high vacuum system}},
volume = {8885},
booktitle = {Laser-Induced Damage in Optical Materials: 2013},
editor = {Gregory J. Exarhos and Vitaly E. Gruzdev and Joseph A. Menapace and Detlev Ristau and MJ Soileau},
organization = {International Society for Optics and Photonics},
publisher = {SPIE},
pages = {88852E},
year = {2013},
doi = {10.1117/12.2047327},
URL = {https://doi.org/10.1117/12.2047327}
}

@techreport{moscatello_phd,
author = "{A. Moscatello}",
type = "{PhD thesis}",
year = {2026},
title = {Stray Light from Particle Contamination in the Virgo and Einstein Telescope Gravitational Waves Detectors},
institution = {University of Padua},
url = {https://hdl.handle.net/20.500.14242/358094}
}

@article{Pierre_Drossart_1990, 
title={A statistical model for the scattering by irregular particles}, 
volume={361}, 
ISSN={0004-637X}, 
DOI={10.1086/185819}, 
journal={The Astrophysical Journal}, 
publisher={American Astronomical Society}, 
author={Pierre Drossart}, 
year={1990} }

@article{Li:26,
author = {Cunhui Li and Zecheng Cui and Jiajie Wang and Xu Tan and Meng Chen},
journal = {Appl. Opt.},
number = {8},
pages = {2416--2424},
publisher = {Optica Publishing Group},
title = {Optical response of a single non-spherical particle and a correction scheme for the dust detection system onboard Tianwen-2},
volume = {65},
month = {Mar},
year = {2026},
url = {https://opg.optica.org/ao/abstract.cfm?URI=ao-65-8-2416},
doi = {10.1364/AO.585904}
}

@Article{amt-7-931-2014,
AUTHOR = {Lurton, T. and Renard, J.-B. and Vignelles, D. and Jeannot, M. and Akiki, R. and Mineau, J.-L. and Tonnelier, T.},
TITLE = {Light scattering at small angles by atmospheric irregular particles: modelling and laboratory measurements},
JOURNAL = {Atmospheric Measurement Techniques},
VOLUME = {7},
YEAR = {2014},
NUMBER = {4},
PAGES = {931--939},
URL = {https://amt.copernicus.org/articles/7/931/2014/},
DOI = {10.5194/amt-7-931-2014}
}

@misc{nandi2026scatteredlightnoiseligo,
      title={Scattered light noise at LIGO Livingston Observatory during O4}, 
      author={Debasmita Nandi and Anamaria Effler and Siddharth Soni and Tabata Aira Ferreira and Robert Schofield and Huyen Pham and Timothy O'Hanlon and V. V. Frolov and Gabriela González},
      year={2026},
      eprint={2605.14143},
      archivePrefix={arXiv},
      primaryClass={astro-ph.IM},
      url={https://arxiv.org/abs/2605.14143}, 
}

\end{document}